\documentclass[12pt,a4paper,endfloat,revised]{article}

\usepackage[utf8]{inputenc}
\usepackage[english]{babel}

\usepackage{amsmath,amssymb}
\usepackage{mathtools}
\usepackage{mathptmx}
\usepackage{upgreek}

\usepackage{graphicx}
\usepackage{textcomp}
\usepackage{gensymb}
\usepackage{siunitx}
\usepackage{booktabs}
\usepackage{makecell}
\usepackage{multirow}
\usepackage{multicol}
\usepackage{caption}
\usepackage{subcaption}
\usepackage{float}
\usepackage{tikz}

\usepackage{appendix}
\usepackage{authblk}
\usepackage{xcolor}
\usepackage{indentfirst}
\usepackage[normalem]{ulem}

\usepackage[a4paper, total={7in, 9in}]{geometry}
\usepackage{setspace}
\usepackage{fancyhdr}
\usepackage{natbib}

\usepackage{blindtext}

\usepackage{hyperref}

\usepackage{cleveref}

\DeclareMathAlphabet{\mathcal}{OMS}{cmsy}{m}{n}

\newif\iftrackchanges
\trackchangestrue

\iftrackchanges
  \newcommand{\new}[1]{\textcolor{black}{#1}}
  \newcommand{\old}[1]{\textcolor{red}{\sout{#1}}}
\else
  \newcommand{\new}[1]{#1}
  \newcommand{\old}[1]{}
\fi

\makeatletter
\newcounter{segappendix}
\renewcommand{\thesegappendix}{\Alph{segappendix}}

\newcommand{\segappend}[2]{%
  \refstepcounter{segappendix}%
  \section*{Appendix \thesegappendix: #1}%
  \addcontentsline{toc}{section}{Appendix \thesegappendix: #1}%
  \protected@edef\@currentlabel{\thesegappendix}%
  \def\@currentlabelname{#1}%
  \label{#2}%
}
\makeatother

\title{\MakeUppercase{Seismic amplitude-variation-with-offset inversion using an algebraically equivalent formulation of the exact Zoeppritz equations and a matrix-free adjoint-state framework}}
\author[1]{Wiktor Waldemar Weibull} \author[1]{Nisar Ahmed}

\affil[1]{Department of Energy and Petroleum Engineering, Faculty of Science and Technology, University of Stavanger, 4036 Stavanger, Norway}

\affil{Corresponding author: nisar.ahmed@uis.no}

\date{03 March 2025}

\makeatletter
\renewcommand{\maketitle}{
  \begin{center}
    \hrule height 0.8pt
    \vspace{0.5em}
    {\LARGE\bfseries \@title \par}
    \vspace{0.5em}
    \hrule height 0.8pt
    \vspace{1em}
    {\large \@author \par}
    \vspace{0.5em}
    {\@date \par}
  \end{center}
  \vspace{2em}
}
\makeatother

\begin{document}

\maketitle

\begin{abstract}
The amplitude-variation-with-offset inversion techniques are formulated to estimate elastic properties by fitting modeled seismic responses to observed data. Solving inverse seismic problems requires minimizing a target objective function for which gradient-based methods are frequently adopted. However, the efficiency and accuracy of these methods depend significantly on the approach used to compute the gradient of the target function. \new{This work first derives an algebraically equivalent reformulation of the exact Knott–Zoeppritz PP equations to obtain a differentiable forward model suitable for gradient-based optimization. The resulting formulation is then used to derive an explicit adjoint-state gradient of a convolution-based objective function for discretized multilayer media, providing an efficient framework for exact nonlinear inversion of P- and S-wave velocities and density, including the consistent treatment of post-critical-angle reflections through the complex-valued Zoeppritz response.} The adjoint state-based solution improves computational efficiency by avoiding numerical approximations while maintaining high accuracy in calculating the gradient for seismic inversion. Additionally, using the exact Zoeppritz equation helps overcome the limitations associated with weak elastic property contrasts across subsurface layers. The least squares target function is minimized using a nonlinear limited-memory quasi-Newton algorithm. \new{The inversion methodology is validated using 1D well-log-based and 2D synthetic seismic data at varying signal-to-noise ratios, including a 500-member ensemble test of sensitivity to the starting model and a Marmousi comparison of Zoeppritz versus linearized Aki--Richards reflectivity at high contrast and wide angles.} Then it is applied to a 2D field data set from the Troll oil and gas field in the Norwegian North Sea. The results demonstrate that the proposed inversion framework provides stable and reliable estimates of elastic property models.
\end{abstract}\label{section:1}

\section{Introduction}\label{section:2}

Seismic amplitude-variation-with-offset (AVO) inversion techniques are widely used in exploration geophysics to estimate subsurface elastic properties, including P-wave velocity, S-wave velocity, and density.  These properties are essential for characterizing the mechanical behavior of rocks and can provide indirect insights into lithology and fluid content when integrated with rock physics models \citep{liu2018stochastic, de2019multimodal, ahmed2024constrained, kjonsberg2024bayesian}. The inversion process integrates seismic observations, well-log, and core measurements, guided by geophysical models and mathematical inversion techniques. Transformation of geophysical data into reservoir properties is an ill-posed inverse problem with a nonunique solution, due to the limited bandwidth of seismic data, noise, and the physical assumptions inherent in forward modeling \citep{tarantola2005inverse, adler2017solving, adler2024deep, ahmed2026enhanced}. The non-uniqueness and instability of seismic inversion problems necessitate the use of robust optimization techniques to explore the model space and identify plausible solutions. In general, there are two main classes of optimization methods \citep{gill2019practical} used in geophysical inversion: global optimization and gradient-based optimization. \par  

In global optimization methods \citep{horst2000introduction, pardalos2013handbook, forst2010optimization, horst2013handbook}, the objective is to search the entire optimal model vector to identify the global minimum of the objective function without premature convergence to local minima. This optimization can be carried out using a variety of approaches, including Monte Carlo sampling methods \citep{FuHu1997}, simulated annealing, genetic algorithms \citep{SenStoffa1995}, linear and nonlinear programming techniques \citep{horst2013handbook}, or hybrid algorithms \citep{SenStoffa1992}. In recent years, global optimization methods have been increasingly applied to solve complex geophysical inverse problems, particularly in seismic amplitude-variation-with-offset inversions \citep{ma2002simultaneous, misra2008global, zhe2013non, fang2017nonlinear, wu2017research, wu2019improved, dupuy2021bayesian, li2021avo, xu2024global}. However, if the problem size is sufficiently small or the computational resources are adequate, global optimization techniques can be highly effective for seismic inversion tasks \citep{SenStoffa}. In practice, global search methods are often computationally intensive and may become impractical for solving large-scale three-dimensional seismic inversion problems due to their high CPU time demand and large search space domains. Furthermore, many global optimization techniques do not offer formal guarantees of reaching the true global optimum. Consequently, gradient-based optimization methods have become the default choice in exploration geophysics, offering faster convergence and lower computational demands \citep{schuster2017seismic}. \par 

Gradient-based optimization techniques are widely used to solve seismic amplitude-variation-with-offset inverse problems \citep{luo2020joint, huang2021directional, wang2022pre, ahmed2022constrained, ahmed2023frequency, wang2023three, correa2025rock, wu2025prestack} due to their computational efficiency and scalability, especially when dealing with large datasets and high-dimensional model spaces. When compared with nonlinear global optimization, gradient-informed methods require much fewer iterations to minimize the mismatch between observed and modeled seismic responses. The gradient provides the steepest descent direction, allowing fast convergence toward a local minimum. However, these methods require the computation of the sensitivity matrix or Jacobian, which links model properties such as P-wave velocity, S-wave velocity, and density to changes in seismic amplitude. \citet{gholami2018constrained} proposed a nonlinear Zoeppritz-based AVO inversion method that combines Levenberg and split Bregman optimization, using exact forward modeling and a linearized Jacobian approximation. \citet{luo2020joint2} adopted a gradient-based AVO inversion technique using the difference-of-convex algorithm combined with proximal operators and a fast iterative shrinkage-thresholding algorithm. This method uses subgradients, and soft-thresholding typically results in slower convergence when dealing with large or ill-posed inverse problems compared to more advanced optimization methods. \citet{huang2021directional} uses Fréchet derivative to compute the gradients of the objective function and applies the split Bregman algorithm \citep{GoldsteinOsher2009} to minimize the regularized data misfit. \citet{correa2025rock} computed the Jacobian based on the linearized Aki and Richards approximation, integrated with rock physics models for the inversion of petrophysical properties, and used the Levenberg-Marquardt algorithm to solve the corresponding nonlinear inverse problems. \citet{wu2025prestack} performed pre-stack inversion using a well-controlled AVO operator, where gradients of the objective function are computed analytically based on a reformulated forward model, and the inversion is solved using projected gradient descent (PGD) and iterative weighted least squares (IRLS). However, this method heavily relies on well-controlled AVO operators and requires an orthogonal experimental design to optimize multiple regularization factors. \citet{ahmed2022constrained} proposed a non-linear seismic AVO inversion method that employs the adjoint state method for efficient gradient computation of P-wave velocity, S-wave velocity, and density. These adjoint-state-based gradients are further extended by \citet{ahmed2024constrained, ahmed2024viscoelastic} to account for reservoir dynamic changes, i.e., saturation and pressure gradients, by integrating with elastic and viscoelastic rock physics models. Their approach combines a convolutional forward model with a linearized approximation of the Zoeppritz equation, which inherently limits its accuracy in cases involving strong elastic contrasts, wide-angle reflections, or highly nonlinear seismic responses.  \par 

\new{To overcome the limitations associated with linearized AVO forward modeling, numerous studies have explored prestack inversion using the exact Zoeppritz equations. \citet{zhi2016amplitude} developed a nonlinear AVO inversion framework based on the exact Zoeppritz equations, in which analytical Fréchet derivatives were incorporated within an iteratively regularized Levenberg--Marquardt algorithm for multilayer elastic parameter estimation. \citet{liu2019accurate} derived analytical Jacobian matrices based on the exact Zoeppritz equations for the direct inversion of dry-rock elastic moduli, while \citet{chai2020elastic} derived exact Jacobian and Hessian matrices for nonlinear inversion of P- and S-wave velocities and density. Extensions to anisotropic inversion were proposed by \citet{Bao2021}, who derived analytical derivatives of exact Zoeppritz-based reflection coefficients and incorporated PP and PS data within a Marquardt optimization framework. More recently, \citet{xu2023prestack} combined exact Zoeppritz inversion with regularization and edge-preserving filtering to improve the stability and structural continuity of the inverted models. Although the above-mentioned studies have derived analytical Jacobian or Fréchet-derivative expressions for the exact Zoeppritz equations. However, these approaches rely on the explicit derivation and assembly of sensitivity matrices, which become increasingly cumbersome as the number of model parameters grows. In contrast, the adjoint-state formulation \citep{plessix2006review, ahmed2023adjoint} computes the gradient of the objective function through a single forward evaluation and a corresponding adjoint recursion, avoiding explicit Jacobian construction. The computational cost of the gradient evaluation is therefore largely independent of the number of model parameters, making the approach particularly attractive for large-scale multilayer inversion problems. In addition, the matrix-free formulation reduces memory requirements and integrates naturally with quasi-Newton optimization algorithms such as L-BFGS. Furthermore, existing exact-Zoeppritz inversion frameworks \citep{liu2019accurate, chai2020elastic, Bao2021, xu2023prestack} are based on real-valued sensitivity formulations, whereas the treatment of complex-valued reflectivities arising beyond the critical angle remains largely unexplored in gradient-based inversion.} \par 

\new{To overcome the limitations associated with linearized AVO approximations, explicit Jacobian-based sensitivity formulations, and the incomplete treatment of complex-valued Zoeppritz responses at post-critical angles, we develop a nonlinear amplitude-versus-offset inversion framework based on the exact Zoeppritz equations \citep{zoeppritz1919viib}. This study presents a unified framework for exact nonlinear prestack seismic inversion based on the full Zoeppritz equations, formulated within the convolutional AVO framework described in the next section. First, the classical Knott--Zoeppritz PP system is reformulated into an algebraically equivalent contrast-based representation, yielding a compact computational graph for the forward and adjoint reflectivity operators and enabling efficient analytical differentiation while preserving the exact physics of the Zoeppritz equation~\citep{lavaud1999pushing, chavent2010nonlinear}, derived in Appendix~\ref{app:equivalent_zoep}. Second, the adjoint-state accelerated analytical gradients with respect to P-wave velocity, S-wave velocity, and density are then derived for discretized multilayer media using an algebraically equivalent Zoeppritz formulation \citep{weibull2026adjoint}. This enables matrix-free gradient computation without explicit Jacobian assembly. The resulting gradients are integrated within a gradient-based optimization algorithm to enable computationally efficient inversion.  Although adjoint-state formulations of the Zoeppritz equations have been introduced previously \citep[e.g.,][]{chavent2010nonlinear}, his work was primarily focused on the two-layer reflectivity problem. Its extension to discretized multilayer media, together with the derivation of analytical adjoint-state gradients and implementation within a nonlinear inversion framework, has not been addressed.} In attempting this extension, we identified that certain relations in the two-layer adjoint system (such as the condition $\lambda_{3} - \lambda_{8}=0$) do not generalize consistently across multiple interfaces, motivating a \new{consistent multilayer derivation}. \new{In addition, the derived forward and adjoint operators consistently account for the complex-valued Zoeppritz response beyond the critical angle, allowing amplitude and phase effects to be incorporated within the inversion and extending its applicability to wide-angle data. Building on the forward and adjoint formulations described above, the inversion is formulated as a constrained nonlinear optimization problem with a regularized $L_2$ least-squares objective function. Model parameters are represented through a bounded sigmoid transform, which imposes physical constraints on P- and S-wave velocity and density, while preserving differentiability of the optimization problem. The resulting optimization problem is solved using a limited-memory quasi-Newton algorithm that efficiently approximates the inverse Hessian without explicit computation or storage \citep{nocedal2006numerical}.} 
\new{The main contributions of this study are:
1. an algebraically equivalent contrast-based reformulation of the exact PP Zoeppritz coefficient suitable for discrete adjoint differentiation;
2. a matrix-free adjoint-state gradient for convolutional multilayer AVO inversion; and
3. a consistent treatment of complex-valued post-critical Zoeppritz reflectivities in a real-valued seismic objective function.
The inversion is validated on 1D well-log tests, a 500-member initial model ensemble, a 2D Marmousi model, a high-contrast Zoeppritz/Aki–Richards comparison, and Troll field data, located in the Norwegian North Sea.}

This work is organized as follows: First, we begin with the seismic inversion method adopted in this study. Then, we describe the applications of the proposed inversion to synthetic and field data. Finally, the key findings are discussed and summarized.

\section{Theory and Method}

\subsection{Convolutional model}
The seismic partially stacked data are modeled by convolving a band-limited source wavelet with the Earth’s angle-dependent PP reflectivity. In practice, both the data and model are sampled, and we work with a discrete forward model. Let $i=0,\dots,N_s-1$ index time/depth samples (interfaces), let $j=0,\dots,N_a-1$ index incidence angles, and let $m=0,\dots,M-1$ index samples of a (possibly angle-dependent) wavelet of length $M$. The discrete forward model is
\begin{equation}\label{eqn:1}
    d^\text{mod}_{i,j} = \sum_{m=0}^{M-1} w_{m,j}\,R_{i-m,j}(\mathbf{m}),
\end{equation}
\new{where $R_{i,j}(\mathbf{m})$ is the PP reflection coefficient at interface/sample $i$ and angle $\theta_j$, computed from the elastic parameters $\mathbf{m}$.}
\new{Equation~}\eqref{eqn:1} \new{ defines the noise-free forward map used in $J_{\mathrm{data}}$; random noise is added only when generating synthetic examples (Section~Applications), not in the inversion loop.}
\new{When $R_{i,j}$ is complex (post-critical incidence), the wavelet $w_{m,j}$ remains real and we set $d^\text{mod}_{i,j}=\operatorname{Re}\{\sum_m w_{m,j}\,R_{i-m,j}(\mathbf{m})\}$ so that $d^\text{mod}_{i,j}$ matches the real-valued partial-angle stacks. Because $w_{m,j}$ is real, this is equivalent to convolving with $\operatorname{Re}(R_{i,j})$; the modeled trace is therefore tied to the in-phase component $\operatorname{Re}(R_{i,j})=|R_{i,j}|\cos\phi_{i,j}$ rather than to $\operatorname{Im}(R_{i,j})$, i.e.\ to signed real reflectivity amplitudes in the usual AVO sense rather than to a separately phase-rotated post-critical wavelet.}

The variation in reflection amplitude with incidence angle is described by the Zoeppritz equations \citep{zoeppritz1919viib}. 
In this study, we focus exclusively on the PP reflection coefficient $R_{PP}$ as a function of incidence angle and elastic properties across an interface, 
which is then convolved with the corresponding angle-dependent source wavelet to model the partial angle stacks.

\new{This formulation is a deliberate idealization of prestack AVO analysis. In field applications, seismic data are normally conditioned to common-reflection-point (or equivalent) gathers of primary PP reflections and corrected for geometrical spreading, transmission losses, and intrinsic attenuation before AVO inversion. Equation~}\eqref{eqn:1} \new{ represents the corresponding one-dimensional modeling step: convolution of a (possibly angle-dependent) source wavelet with interface reflectivity sampled at discrete incidence angles. The forward operator therefore targets primary PP partial-angle stacks rather than full two- or three-dimensional wave-equation modeling. It assumes (i)~primary PP reflections in the angle stacks, (ii)~a one-dimensional layered elastic medium along each trace, (iii)~amplitude variations dominated by interface contrasts through $R_{PP}(\theta)$, and (iv)~negligible interbed multiples and converted-wave contamination in the stacks. These assumptions are standard in gradient-based AVO inversion because they isolate the sensitivity of partial stacks to elastic parameters while keeping the forward map differentiable for adjoint-state gradient computation.}

\subsection{Seismic inversion algorithm}

To solve the seismic amplitude-variation-with-offset inverse problem, we formulate it as a nonlinear optimization problem. 

We denote by $\textbf{m} = [v_p,\: v_s,\: \rho]^T)$ the physical elastic model and by $\textbf{z}$ the unconstrained optimization variable. The bounded physical parameters are obtained through the mapping $\textbf{m} = m(\textbf{z})$. The inversion problem is therefore written as
\begin{equation}
    \textbf{z}^{\ast} = \text{arg min} \:\:\: J_{\mathrm{tot}}(m(\textbf{z})),
\end{equation}
\new{The data misfit $J_{\mathrm{data}}$ is the least-squares ($L_{2}$-norm) objective represented in discretized form by}
\begin{equation}\label{eq:Jdata}
J_{\mathrm{data}}(\mathbf{m}) = \frac{1}{2}\sum_{j=0}^{N_a-1}\sum_{i=0}^{N_s-1}
\left(d^\text{obs}_{i,j} - \sum_{m=0}^{M-1} w_{m,j}\,R_{i-m,j}(\mathbf{m}) \right)^2,
\end{equation}
where $\mathbf{m}=\{v_{p,i},v_{s,i},\rho_i\}_{i=0}^{N_s-1}$ denotes the physical elastic model.
Here $d^\text{obs}_{i,j}$ are the observed partial angle stacks, and the predicted data are obtained by the discrete convolution in equation~\eqref{eqn:1}.
\new{The total objective minimized in the examples is $J_{\mathrm{tot}}=J_{\mathrm{data}}+J_{\mathrm{Tikh}}+J_{\mathrm{TV}}$ (Section~}\nameref{sec:constraints_reg}\new{).}
This discretization is essential for gradient-based inversion: the adjoint-state method used below computes exact derivatives of the discrete computational graph mapping $\mathbf{m}\mapsto R_{i,j}(\mathbf{m})\mapsto d^\text{mod}_{i,j}\mapsto J_{\mathrm{data}}(\mathbf{m})$.

There are numerous optimization techniques to minimize the objective function. We applied the limited-memory quasi-Newton algorithm \citep{nocedal2006numerical} to solve the objective function and the iterative update of the model. The iterative equation is
\begin{equation}\label{eqn:5}
    \textbf{x}^{(k+1)} = \textbf{x}^{(k)} - \gamma^{(k)} \: \textbf{H}^{(k)^{-1}} \nabla\textbf{ J}^{(k)},
\end{equation}
where $k$ is the iteration index, $\gamma$ denotes step length, and $\textbf{H}^{(k)^{-1}}$ indicates the approximate Hessian inverse matrix at a given iteration index. 
\new{Gradients of $J_{\mathrm{data}}$ with respect to physical parameters are computed using the adjoint-state method;} which efficiently computes the derivatives of the reflection coefficients with respect to the physical parameters. 
In the remainder of this section, we first define the discretized forward model used to compute the angle-dependent PP reflectivities $R_{ij}$ and the corresponding modeled data, then present the adjoint-state recursion for computing sensitivities of $R_{ij}$, and finally assemble the gradients with respect to the physical parameters $(v_p, v_s, \rho)$ used by the optimizer.

\subsection{Discretized forward modelling}

We consider a 1D layered elastic medium parameterized by P-wave velocity $v_p(z)$, S-wave velocity $v_s(z)$ and density $\rho(z)$, sampled on a regular grid $z_i$, $i=0,\dots,N_s-1$. We denote the sampled parameters as
\[
 v_{p,i},\ v_{s,i},\ \rho_i, \qquad i=0,\dots,N_s-1.
\]

Let $\theta_j$, $j=0,\dots,N_a-1$ be the set of incidence angles. At each interface at depth index $i$, we use the parameter values at depth indices $i-1$, $i$, and $i+1$ to represent the values above, at, and below the interface, respectively. We denote the PP reflection coefficient at interface $i$ and angle $\theta_j$ by $R_{ij}$. \new{The reflection coefficients form a matrix $R$ with entries in $\mathbb{C}$ (reducing to $\mathbb{R}$ precritically).}


\paragraph{Interface parameters}


\new{At each interface, the PP reflection coefficient is computed from an algebraically equivalent reformulation of the exact Knott--Zoeppritz equations }\citep{lavaud1999pushing,chavent2010nonlinear}\new{. Appendix~}\ref{app:equivalent_zoep}\new{ derives this form from the classical system; the sequence below is the multilayer discretization used in the forward model and in the adjoint-state recursion of Appendix~}\ref{app:adjoint_details}\new{. The reformulation maps reflectivity to a fixed computational graph of contrasts, square roots, and rational operations, which simplifies exact differentiation relative to the trigonometric and slowness-dependent classical expressions.}

At each interface $(i,j)$, the following non-dimensional contrasts and auxiliary parameters are defined:
\begin{align}\label{eq:interface_params}
 e_{r,ij} &= \frac{\rho_{i+1} - \rho_i}{\rho_{i+1} + \rho_i}, \\
 e_{p,ij} &= \frac{v_{p,i+1}^2 - v_{p,i}^2}{v_{p,i+1}^2 + v_{p,i}^2}, \\
 e_{s,ij} &= \frac{v_{s,i+1}^2 - v_{s,i}^2}{v_{s,i+1}^2 + v_{s,i}^2}, \\
 \chi_{ij} &= \frac{v_{s,i+1}^2 + v_{s,i}^2}{v_{p,i+1}^2 + v_{p,i}^2}.
\end{align}
The sequence of intermediate quantities is
\begin{align}
 e_{ij} &= e_{s,ij} + e_{r,ij}, \\
 f_{ij} &= 1 - e_{r,ij}^2, \\
 S_{1,ij} &= \frac{\chi_{ij}}{1 - e_{p,ij}}, \\
 S_{2,ij} &= \frac{\chi_{ij}}{1 + e_{p,ij}}, \\
 T_{1,ij} &= \frac{1}{1 - e_{s,ij}}, \\
 T_{2,ij} &= \frac{1}{1 + e_{s,ij}}, \\
 q^2_{ij} &= S_{1,ij} \sin^2 \theta_{j}, \\
 M_{1,ij} &= \sqrt{S_{1,ij} - q^2_{ij}}, \\
 M_{2,ij} &= \sqrt{S_{2,ij} - q^2_{ij}}, \\
 N_{1,ij} &= \sqrt{T_{1,ij} - q^2_{ij}}, \\
 N_{2,ij} &= \sqrt{T_{2,ij} - q^2_{ij}}, \\
 D_{ij} &= e_{ij} q^2_{ij}, \\
 A_{ij} &= e_{r,ij} - 2D_{ij}, \\
 K_{ij} &= 2D_{ij} - A_{ij}, \\
 B_{ij} &= 1 - K_{ij}, \\
 C_{ij} &= 1 + K_{ij}.
\end{align}
The terms $P_{ij}$ and $Q_{ij}$ are then defined as
\begin{align}
 P_{ij} &= M_{1,ij}\left(B_{ij}^2 N_{1,ij} + f_{ij} N_{2,ij}\right)
 + 16 e_{ij} D_{ij} M_{1,ij} M_{2,ij} N_{1,ij} N_{2,ij}, \\
 Q_{ij} &= M_{2,ij}\left(C_{ij}^2 N_{2,ij} + f_{ij} N_{1,ij}\right)
 + 4 q^2_{ij} A_{ij}^2.
\end{align}
Finally, the PP reflection coefficient is
\begin{equation}\label{eq:reflection_coeff}
 R_{ij} = \frac{P_{ij} - Q_{ij}}{P_{ij} + Q_{ij}}.
\end{equation}
\new{Equations~}\eqref{eq:interface_params}\new{--}\eqref{eq:reflection_coeff} \new{are algebraically equivalent to the classical Knott--Zoeppritz PP system when the contrasts are defined as in Appendix~}\ref{app:equivalent_zoep}\new{. We verified numerical agreement between the classical and reformulated expressions to machine precision for representative elastic models and incidence angles (precritical and post-critical).}

\paragraph{Post-critical reflectivity}
\new{For interfaces with $v_{p,i+1} > v_{p,i}$, the critical angle is $\theta_{c,ij}=\arcsin(v_{p,i}/v_{p,i+1})$. For $\theta_j<\theta_{c,ij}$ all radicands in equations~\eqref{eq:interface_params}--\eqref{eq:reflection_coeff}
are positive and $R_{ij}$ is real. At $\theta_j\ge\theta_{c,ij}$ the radicand $S_{2,ij}-q_{ij}^2$, which corresponds to the vertical slowness of the transmitted P~wave, becomes negative ($T_{1,ij}-q_{ij}^2$ and $T_{2,ij}-q_{ij}^2$ may likewise become negative when a shear velocity exceeds $v_{p,i}$, whereas $M_{1,ij}\propto\cos\theta_j$ remains real). Rather than
clipping the incidence angles at $\theta_{c,ij}$, we evaluate $M_{k,ij}$ and $N_{k,ij}$ as principal complex square roots (argument in $([-\pi,\pi])$, so that a negative radicand yields a purely imaginary vertical slowness with
non-negative imaginary part. Physically, this is the evanescent continuation of the transmitted wavefield: beyond the critical angle the transmitted wave becomes inhomogeneous and decays exponentially away from the interface, and the chosen branch enforces this boundedness (the radiation condition). The reflection coefficient $R_{ij}$ then becomes complex, with its modulus and argument describing the amplitude and phase rotation of the post-critical reflection, and equation~\eqref{eq:reflection_coeff} remains valid and continuous through the critical angle without any special-case treatment.
The complex $R_{ij}$ is mapped to a real trace through $\operatorname{Re}(R_{ij})$ in the real-wavelet convolution of equation~\eqref{eqn:1}. This wide-angle response is precisely where the linearized Aki--Richards approximation breaks down, and where sensitivity to the velocity contrast across the interface is greatest. We note that, beyond critical incidence, the Knott--Zoeppritz plane-wave system remains approximate for point-source data because it excludes head waves and interference effects \citep{aki2002quantitative}. In the discussion section, we compare this exact nonlinear response with the Aki--Richards approximation at wide angles for the high-contrast two-layer test of Section~\nameref{sec:zoep_aki}.}

\paragraph{Reproduction recipe}
\new{The forward map and gradients are reproduced as follows for each interface $(i,j)$ and prescribed angles $\{\theta_j\}$:}
\begin{enumerate}
 \item\new{Specify $(v_{p,i},v_{s,i},\rho_i)$ and form $e_{r,ij}$, $e_{p,ij}$, $e_{s,ij}$, and $\chi_{ij}$ from Equations~}\eqref{eq:interface_params}\new{.}
 \item\new{Compute $e_{ij}$, $f_{ij}$, $S_{1,ij}$, $S_{2,ij}$, $T_{1,ij}$, $T_{2,ij}$, and $q_{ij}^2=S_{1,ij}\sin^2\theta_j$.}
 \item\new{Set $M_{k,ij}=\sqrt{S_{k,ij}-q_{ij}^2}$ and $N_{k,ij}=\sqrt{T_{k,ij}-q_{ij}^2}$, using principal complex square roots when a radicand is negative.}
 \item\new{Form $D_{ij}$, $A_{ij}$, $K_{ij}$, $B_{ij}$, $C_{ij}$, then $P_{ij}$, $Q_{ij}$, and $R_{ij}=(P_{ij}-Q_{ij})/(P_{ij}+Q_{ij})$.}
 \item\new{Build synthetic angle gathers with a real band-limited wavelet: $d_{i,j}^\text{mod}=\operatorname{Re}\{\sum_m w_{m,j}R_{i-m,j}\}=\sum_m w_{m,j}\operatorname{Re}(R_{i-m,j})$ (Equation~}\eqref{eqn:1}\new{).}
 \item\new{For inversion, compute $\partial J_{\mathrm{data}}/\partial R_{ij}$ from equation~}\eqref{eq:Jdata}\new{, obtain $\partial R_{ij}/\partial e_{\cdot,ij}$ and $\partial R_{ij}/\partial\chi_{ij}$ by the adjoint-state recursion in Appendix~}\ref{app:adjoint_details}\new{ (complex arithmetic when $R_{ij}$ is complex), map to $\partial J_{\mathrm{data}}/\partial v_{p,i}$, $\partial J_{\mathrm{data}}/\partial v_{s,i}$, and $\partial J_{\mathrm{data}}/\partial\rho_i$ via Appendix~}\ref{app:expanded_gradients}\new{, and retain the real part of each term because $J_{\mathrm{data}}$ is real-valued.}
\end{enumerate}
\new{Steps 1--4 define reflectivity; step~5 is the convolutional forward model; step~6 is the gradient in update~}\eqref{eqn:5}\new{. The discrete convolution may be evaluated by direct summation or by any equivalent implementation; forward modeling and adjoint residual projection must use matched discrete operators.}
\new{The reflectivity derivatives in step~6 are evaluated in complex arithmetic when $R_{ij}$ is complex, and the real-valued objective retains only the part consistent with the $\operatorname{Re}(R_{ij})$ entering $d^\text{mod}$.}

\subsection{Adjoint-state formulation}

To compute the gradient $\mathbf{g}$ required for the optimization algorithm in equation~\eqref{eqn:5}, 
we need the derivatives of the reflection coefficients $R_{ij}$ with respect to the physical parameters. 
The adjoint-state method provides an efficient way to compute these derivatives \citep{ahmed2023adjoint}.
\new{The reverse sweep is applied to the reformulated reflectivity graph of Appendix~}\ref{app:equivalent_zoep}\new{.}
We are interested in the sensitivity of $R_{ij}$ 
with respect to the intermediate parameters $e_{r,ij}$, $e_{p,ij}$, $e_{s,ij}$ and $\chi_{ij}$. The derivation can be understood as follows.
For each pair $(i,j)$, the computation of $R_{ij}$ from the primary variables $(e_{r,ij}, e_{p,ij}, e_{s,ij}, \chi_{ij})$ 
proceeds through a sequence of intermediate states defined in equations~\eqref{eq:interface_params}--\eqref{eq:reflection_coeff}. 

This computational sequence maps the primary variables to the final reflection coefficient:
\begin{multline*}
 (e_{r,ij}, e_{p,ij}, e_{s,ij}, \chi_{ij}, \theta_{j})
 \mapsto
 (e_{ij}, f_{ij}, S_{1,ij}, S_{2,ij}, T_{1,ij}, T_{2,ij}, q^2_{ij}, M_{1,ij}, M_{2,ij}, \\
 N_{1,ij}, N_{2,ij}, D_{ij}, A_{ij}, K_{ij}, B_{ij}, C_{ij}, P_{ij}, Q_{ij}, R_{ij}).
\end{multline*}
The adjoint-state method associates with each intermediate quantity in this sequence an adjoint variable and computes the sensitivities by a reverse sweep through this computational graph. For readability, the full Lagrangian, reverse-sweep recursion, and the resulting sensitivities are provided in Appendix~\ref{app:adjoint_details}. 

\subsection{Gradients with respect to physical parameters}

To compute sensitivities needed for $\partial J_{\mathrm{tot}}/\partial \mathbf{z}$ in equation~\eqref{eqn:5}, we first obtain $\partial J_{\mathrm{data}}/\partial \mathbf{m}$ and apply the chain rule. The derivative of $J_{\mathrm{data}}$ with respect to the reflection coefficient $R_{ij}$ is obtained by differentiating equation~\eqref{eq:Jdata}, giving $\partial J_{\mathrm{data}}/\partial R_{ij}$ (the residual term in the adjoint, not the full gradient). 
The derivative of $R_{ij}$ with respect to the physical parameters is computed via the adjoint-state method; the corresponding Lagrangian-based derivation and reverse-sweep expressions are given in Appendix~\ref{app:adjoint_details}. 
The chain rule then gives:
\[
\frac{\partial J}{\partial x_k} = \sum_{i,j} \frac{\partial J}{\partial R_{ij}} \frac{\partial R_{ij}}{\partial x_k},
\]
where $x_k$ represents any physical parameter ($v_{p,i}$, $v_{s,i}$, or $\rho_i$). 

\paragraph{Derivative of the objective function with respect to reflectivities}

The derivative $\partial J_{\mathrm{data}}/\partial R_{ij}$ is obtained by differentiating equation~\eqref{eq:Jdata} with respect to the reflection coefficient $R_{ij}$:
\begin{equation}
\frac{\partial J_{\mathrm{data}}}{\partial R_{ij}} =
-\sum_{m=0}^{M-1} w_{m,j}\left(d^\text{obs}_{i+m,j} - \sum_{p=0}^{M-1} w_{p,j}\,R_{i+m-p,j}\right),
\end{equation}
where $d^\text{obs}_{i+m,j}$ denotes the observed data at time/depth sample $i+m$ and angle $j$, and $w_{m,j}$ are the (possibly angle-dependent) wavelet coefficients used in equation~\eqref{eqn:1}. This expression is a discrete cross-correlation of the data residuals (the terms in parentheses) with the wavelet. The summation over $m$ accounts for all data samples where the reflectivity sample $R_{ij}$ contributes through the discrete convolution in the forward model.

To obtain gradients with respect to the physical parameters, we apply the chain rule through the definitions of $e_{r,ij}$, $e_{p,ij}$, $e_{s,ij}$ and $\chi_{ij}$. Each parameter at depth index $i$ influences at most two adjacent interfaces, giving rise to two contributions that must be summed.
The full expanded expressions for $\partial J_{\mathrm{data}}/\partial \rho_i$, $\partial J_{\mathrm{data}}/\partial v_{p,i}$, and $\partial J_{\mathrm{data}}/\partial v_{s,i}$ (including the two-interface contributions and the final summation over angles) are provided in Appendix~\ref{app:expanded_gradients}. 

\subsection{Constraints, parameterization, and regularization}\label{sec:constraints_reg}

\new{All inversion examples minimize a total objective
\begin{equation}\label{eq:Jtot}
J_{\mathrm{tot}} = J_{\mathrm{data}} + J_{\mathrm{Tikh}} + J_{\mathrm{TV}},
\end{equation}
where $J_{\mathrm{data}}$ is given by equation~}\eqref{eq:Jdata}\new{. The reference model $\mathbf{m}^{0}=(v_{p}^{0},v_{s}^{0},\rho^{0})$ is the initial low-frequency elastic profile at each depth sample. Regularization is evaluated on the physical parameters $(v_{p,i},v_{s,i},\rho_i)$ after they are recovered from the optimization variables.}

\paragraph{Tikhonov and total-variation penalties}
\new{Stability toward the reference model is imposed by a Tikhonov term
\begin{equation}\label{eq:Jtikh}
J_{\mathrm{Tikh}} = \sum_{k\in\{p,s,\rho\}} \beta_k \sum_{i=0}^{N_s-1} \bigl(m_{k,i}-m_{k,i}^{0}\bigr)^{2},
\end{equation}
where $m_{p,i}=v_{p,i}$, $m_{s,i}=v_{s,i}$, and $m_{\rho,i}=\rho_i$, and $\beta_k\ge 0$ are weights. Lateral/vertical structure is encouraged by a smoothed total-variation penalty applied to the same deviations $\mathbf{d}_k=\mathbf{m}_k-\mathbf{m}_k^{0}$:
\begin{equation}\label{eq:Jtv}
J_{\mathrm{TV}} = \sum_{k\in\{p,s,\rho\}} \alpha_k \sum_{i=0}^{N_s-2} \sqrt{\bigl[(D_f \mathbf{d}_k)_i\bigr]^2 + \varepsilon_k^2},
\end{equation}
with forward difference operator $(D_f\mathbf{d})_i = d_{i+1}-d_i$, weights $\alpha_k\ge 0$, and smoothing parameters $\varepsilon_k>0$. The weights $\alpha_k$, $\beta_k$, and $\varepsilon_k$ are chosen empirically.}

\paragraph{Logistic bound constraints}
\new{Because L-BFGS is unconstrained, each bounded physical parameter $p\in[p_{\min},p_{\max}]$ (e.g.\ $v_p$, $v_s$, $\rho$, or a ratio) is represented by an unconstrained optimization variable $x$ through
\begin{equation}\label{eq:logistic}
p(x) = p_{\min} + \frac{p_{\max}-p_{\min}}{1+\exp(-\kappa x)},
\end{equation}
where $\kappa>0$ sets the transition sharpness. At each iteration, $x$ is mapped to $p(x)$ before forward modeling; gradients with respect to $x$ follow from the logistic chain rule. Bounds $[p_{\min},p_{\max}]$ are taken from well-log information (field) or from prior model envelopes (synthetics).}

\paragraph{$V_S/V_P$ reparameterization}
\new{In Examples~2, 3, 4, and the field case we invert $(v_p,\gamma,\rho)$ with $v_s=\gamma v_p$ and $\gamma$ bounded by the same logistic map ($\gamma_{\min}\le\gamma\le\gamma_{\max}$). Example~1 uses the same $J_{\mathrm{tot}}$ and logistic bounds but update $(v_p,v_s,\rho)$ directly. After forming gradients of $J_{\mathrm{tot}}$ with respect to $(v_p,v_s,\rho)$, the $\gamma$ parameterization applies
\begin{equation}\label{eq:vsvp_chain}
\frac{\partial J}{\partial v_p} \leftarrow \frac{\partial J}{\partial v_p} + \gamma\,\frac{\partial J}{\partial v_s}, \qquad
\frac{\partial J}{\partial \gamma} = v_p\,\frac{\partial J}{\partial v_s},
\end{equation}
before mapping to the optimization variables.}

\subsection{Inversion workflow}
\begin{figure}[H]
    \centering
    \includegraphics[scale=0.9]{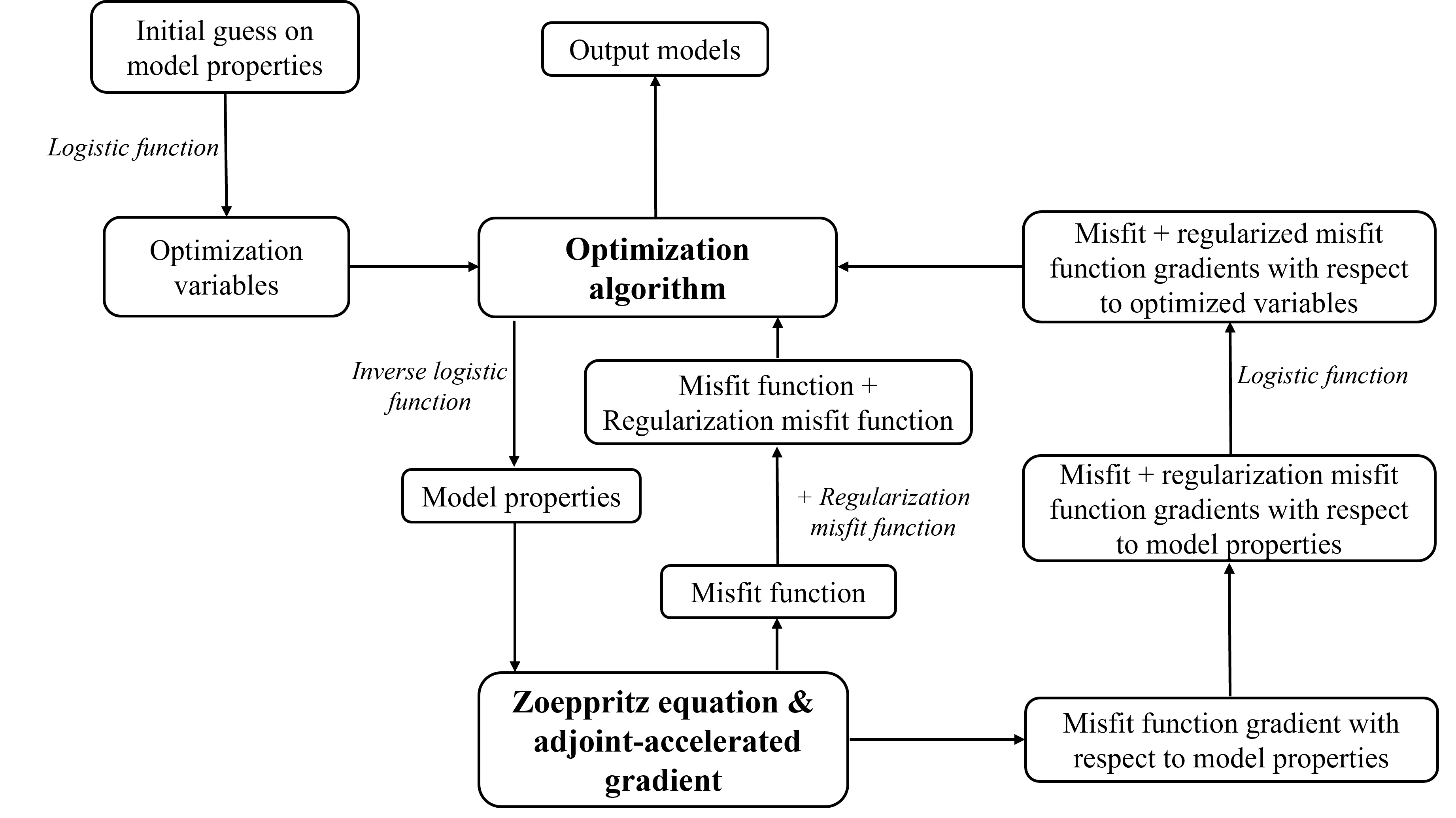}
    \caption{Flow diagram summarizing the stepwise structure of the inversion methodology.}
    \label{fig:1}
\end{figure}

\new{The complete inversion workflow is summarized in Figure}~\ref{fig:1}. 
\new{At each iteration of the limited-memory quasi-Newton optimization, the unconstrained optimization variables are first mapped to bounded physical parameters using the logistic transformation in equation~\eqref{eq:logistic}. 
If the $V_S/V_P$ parameterization is used, the S-wave velocity is then obtained from $v_s=\gamma v_p$. The exact Zoeppritz PP reflection coefficients are computed for all interfaces and incidence angles using equation~\eqref{eq:reflection_coeff}, and the modeled angle stacks are generated by convolution with the corresponding source wavelets. The data-misfit term $J_{\rm data}$ is then evaluated and combined with the Tikhonov and smoothed total-variation penalties to form the total objective function.}
\new{
\begin{equation}
    J_{\rm tot} = J_{\rm data} + J_{\rm Tikh} + J_{\rm TV}.
\end{equation}
The gradient of $J_{\rm data}$ with respect to the physical elastic parameters is computed using the discrete adjoint-state recursion described in Appendix~\ref{app:expanded_gradients}. The gradients of the regularization terms are then added, and the chain rule is applied to account for both the optional $V_S/V_P$ reparameterization and the logistic mapping from unconstrained to bounded variables. The resulting gradient with respect to the optimization variables is used by the L-BFGS algorithm to update the model. This process is repeated until the stopping criterion is satisfied.}
\new{
The resulting algorithm can be written as:
}
\begin{enumerate}
    \item \new{Initialize the physical elastic model and define the reference model 
    $\mathbf{m}_0=(v_{p,0},v_{s,0},\rho_0)$ used in the regularization terms.}

    \item \new{Map the initial bounded physical parameters to unconstrained optimization variables.}

    \item \new{At iteration $k$, map the current optimization variables to bounded physical parameters using the logistic transformation.}

    \item \new{If the ratio parameterization is used, compute 
    $v_s^{(k)}=\gamma^{(k)}v_p^{(k)}$.}

    \item \new{Compute the exact Zoeppritz PP reflectivity 
    $R_{ij}^{(k)}$ for all interfaces and incidence angles.}

    \item \new{Generate the modeled seismic data by convolving 
    $R_{ij}^{(k)}$ with the angle-dependent source wavelets.}

    \item \new{Evaluate the total objective function 
    $J_{\rm tot}^{(k)}=J_{\rm data}^{(k)}+J_{\rm Tikh}^{(k)}+J_{\rm TV}^{(k)}$.}

    \item \new{Compute the adjoint-state gradient of $J_{\rm data}^{(k)}$ with respect to $v_p$, $v_s$, and $\rho$.}

    \item \new{Add the gradients of the Tikhonov and total-variation regularization terms.}

    \item \new{Apply the chain rule for the $V_S/V_P$ parameterization, when used, and for the logistic transformation to obtain the gradient with respect to the unconstrained optimization variables.}

    \item \new{Update the optimization variables using the L-BFGS search direction and line search.}

    \item \new{Repeat steps 3--11 until convergence.}
\end{enumerate}

\section{Applications}

\new{In this section, we evaluate the proposed inversion through (i)~deterministic 1D recovery tests with controlled data noise (Examples~1 and 2), (ii)~an ensemble experiment that quantifies sensitivity to the starting model while holding the data fixed (Example~3), (iii)~a 2D Marmousi test with noise and regularization (Example~4), (iv)~a Zoeppritz versus Aki--Richards comparison against wave-equation reference data on a high-contrast two-layer test, and (v)~a field-data application. All examples minimize $J_{\mathrm{tot}}$ with logistic bounds and Tikhonov plus smoothed total-variation regularization as defined in Section~\nameref{sec:constraints_reg}. Examples~1 inverts $(v_p,v_s,\rho)$ directly; Examples~2, 3, 4, and the Troll field case use the $\gamma=v_s/v_p$ parameterization. Field applications use 2D prestack data from the Troll Field (Norwegian North Sea); bounds follow well-log priors, and regularization weights are tuned empirically.}

\subsection{Synthetic data example}

\begin{figure}[H]
    \centering
    \includegraphics[scale = 0.5]{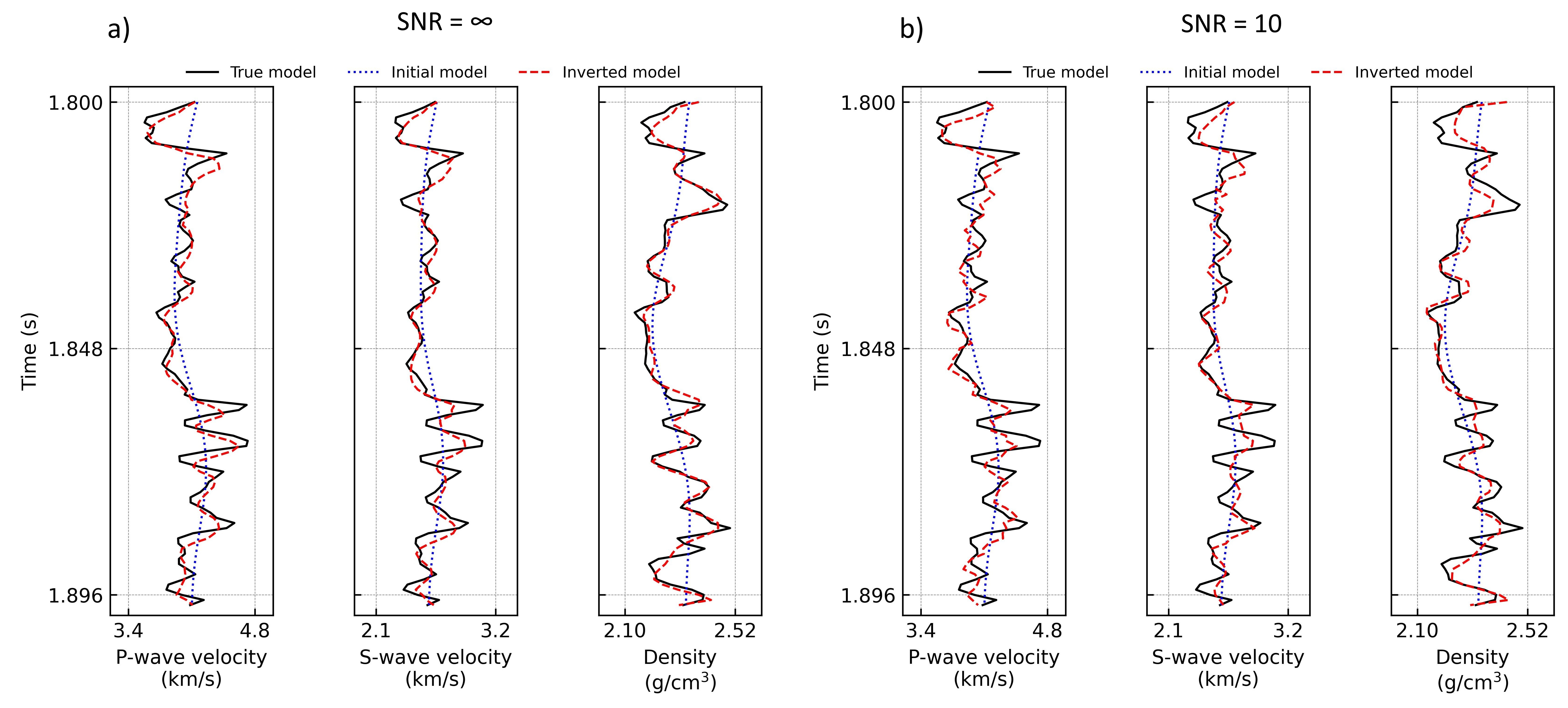}
    \caption{Inverted P- and S-wave velocities and density shown alongside the initial and true models. \new{(a) Noise-free case (SNR = $\infty$ and (b) noisy case (SNR = 10).}}
    \label{fig:2}
\end{figure}

To assess the feasibility and robustness of the proposed amplitude versus offset inversion method, synthetic examples are generated using well-log curves (P- and S-wave velocities and density) from the dataset provided by \citep{grana2021seismic} as model parameters.
\new{Example~1 is a deterministic recovery test: each uses a single smoothed initial model and noise-free and noisy data generated once at a fixed signal-to-noise ratio, thereby assessing data fit and convergence rather than ensemble variability. Synthetic seismic traces up to the maximum incident angle 60$\degree$, with an interval of 5$\degree$, are generated by convolving a 45Hz Ricker wavelet with the P-wave reflectivity calculated using the exact Zoeppritz equation. Zero-mean Gaussian noise was added to the synthetic seismic data. The noise level was controlled through a prescribed signal-to-noise ratio (SNR), with the noise variance scaled relative to the variance of the noise-free data.} To construct the initial model, the true model was smoothed to reduce high-frequency variations. Figure \ref{fig:2}a and \ref{fig:2}b show the inversion results for the elastic properties \new{for the noise-free (SNR = $\infty$) and noisy (SNR = 10) cases, respectively}; red curves indicate the inverted models, the curves in black represent the true models, and the blue dash-dot lines are the initial models. The inverted elastic properties in both panels (a) and (b) generally agree well with the true models. \new{Example~1 inverts $(v_p,v_s,\rho)$ directly under the framework of Section~\nameref{sec:constraints_reg}}.\par 

\new{Table~\ref{tab:1} summarizes performance metrics such as mean absolute error (MAE), mean squared error (MSE), coefficient of determination ($R^2$), and correlation coefficient (r) for noise-free (SNR $=\infty$) and noisy (SNR $=10$) data. In the noise-free case, the proposed inversion framework recovers the elastic parameters with relatively low errors and strong correlations with the reference models, with density exhibiting the highest accuracy. The introduction of noise degrades the inversion performance; however, the predicted models remain reasonably correlated with the true elastic properties.}

\begin{figure}[H]
    \centering
    \includegraphics[scale = 0.6]{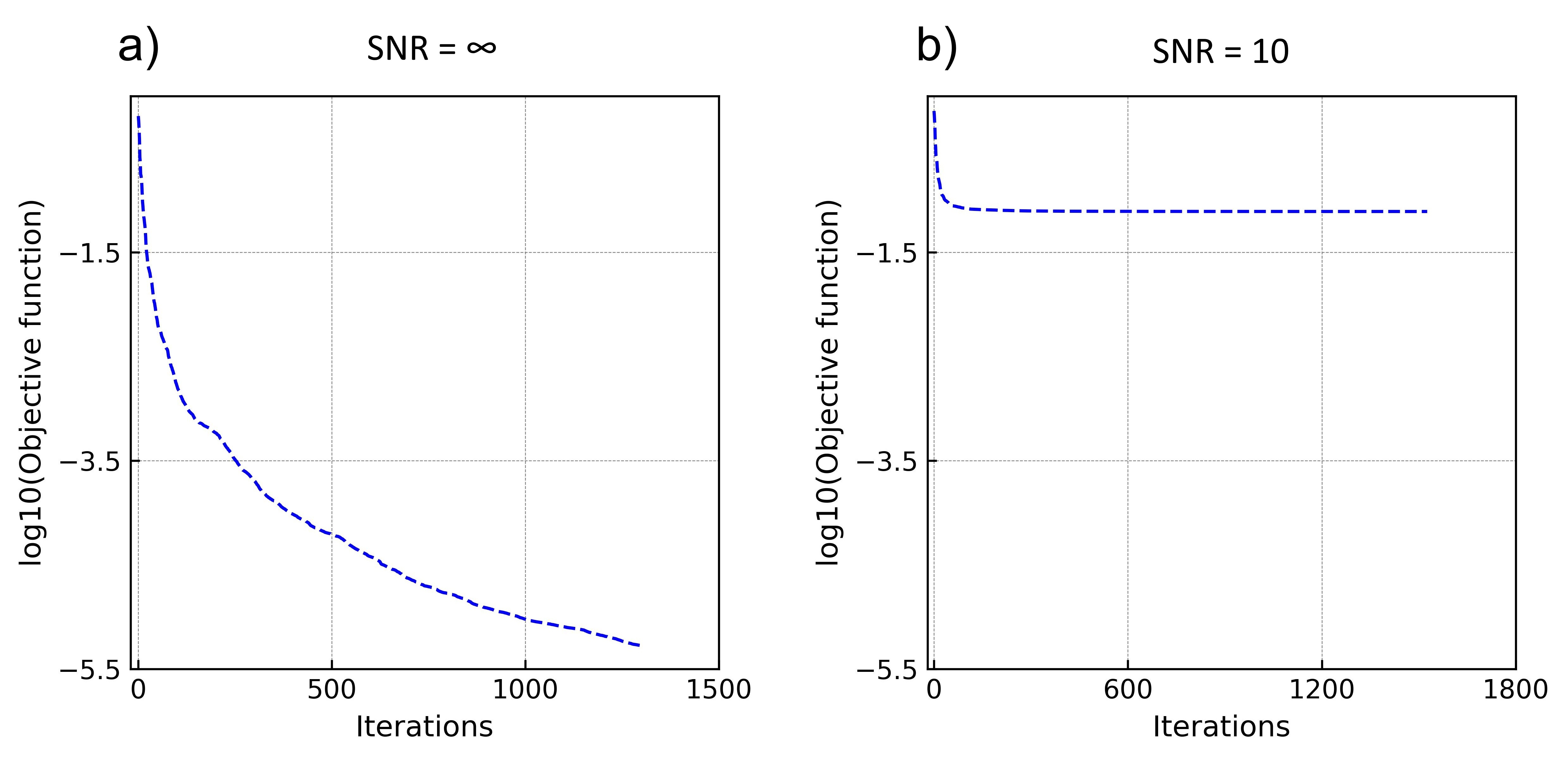}
    \caption{Convergence plot of the inversion in \new{(a) SNR = $\infty$ and (b) SNR = 10 showing $\log_{10}$} of the objective function versus iteration numbers.}
    \label{fig:3}
\end{figure}

The convergence behavior of the inversion method is assessed by plotting $J_{\mathrm{tot}}$ as a function of iteration number (Figures~\ref{fig:3}a and \ref{fig:3}b), on a logarithmic scale, i.e., $\log_{10}(J_{\mathrm{tot}})$. The objective function decreases by approximately five orders of magnitude in the noise-free case. However, the reduction is less pronounced in the noisy case, resulting in a higher converged objective-function value than in the noise-free case. This indicates that the inversion method effectively minimizes the data mismatch and progressively improves the agreement between the observed and predicted seismic data. \par

\begin{table}[ht]
\centering
\caption{\new{Example 1: Inversion performance metrics for P- and S-wave velocities, and density under different noise levels.}}
\label{tab:1}
\begin{tabular}{lcccc}
\hline
Property & MSE & MAE & $R^2$ & $r$ \\
\hline
\multicolumn{5}{c}{SNR = $\infty$} \\
\hline
$V_P$  & 0.0195 & 0.1026 & 0.661 & 0.824 \\
$V_S$  & 0.0093 & 0.0713 & 0.682 & 0.834 \\
$\rho$ & 0.0012 & 0.0256 & 0.852 & 0.927 \\
\hline
\multicolumn{5}{c}{SNR = 10} \\
\hline
$V_P$  & 0.0304 & 0.1432 & 0.471 & 0.700 \\
$V_S$  & 0.0114 & 0.0842 & 0.609 & 0.801 \\
$\rho$ & 0.0024 & 0.0385 & 0.699 & 0.840 \\
\hline
\end{tabular}
\end{table}

\begin{figure}[H]
    \centering
    \includegraphics[scale = 0.52]{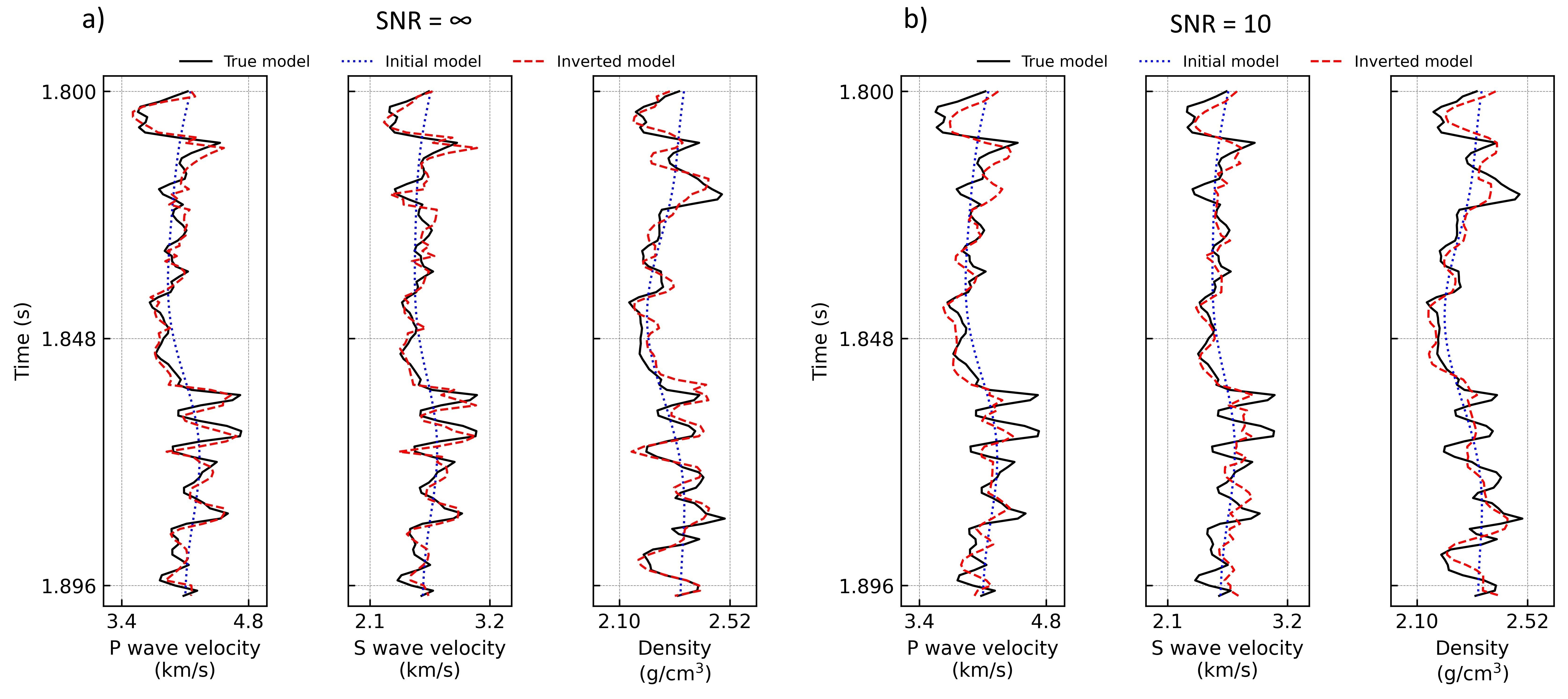}
    \caption{Results of inversion: predicted elastic properties (P- and S-wave velocities and density) compared with the initial and true models. \new{(a) Noise-free case (SNR = $\infty$ and (b) noisy case (SNR = 10).}}
    \label{fig:4}
\end{figure}

\new{In Example~2 (Figure~\ref{fig:4}), $v_p$ and $\rho$ are updated jointly with the ratio $\gamma=v_s/v_p$ via the parameterization in Section~\nameref{sec:constraints_reg} ($v_s=\gamma v_p$, with $\gamma$ logistic-bounded), which stabilizes $v_s$ when the ratio is confined to sedimentary-rock ranges (e.g.\ $\sim$0.4--0.6). For consistency with the previous example, the same forward-modeling setup, incident-angle range ($0^\circ$--$60^\circ$ with $5^\circ$ increments), and initial model were employed. The inversion was performed for both a noise-free case ($\mathrm{SNR}=\infty$) and a noisy case ($\mathrm{SNR}=10$).} The inverted models (Figure \ref{fig:4}a and \ref{fig:4}b) show a good correlation with the true reference models, demonstrating the capability of the proposed inversion method to recover the elastic properties. The predicted models capture the main spatial trends and reservoir characteristics (1.8–1.85 s) in both plots, demonstrating the reliability of the proposed method. \new{Table \ref{tab:2} shows the performance metrics such as MSE, MAE, $R^2$ and $r$ for noise-free ($SNR =\infty$) and noisy ($SNR=10$) data. Under both noise conditions, the method predicts model properties with lower errors and higher correlation metrics.}

\begin{figure}[H]
    \centering
    \includegraphics[scale = 0.6]{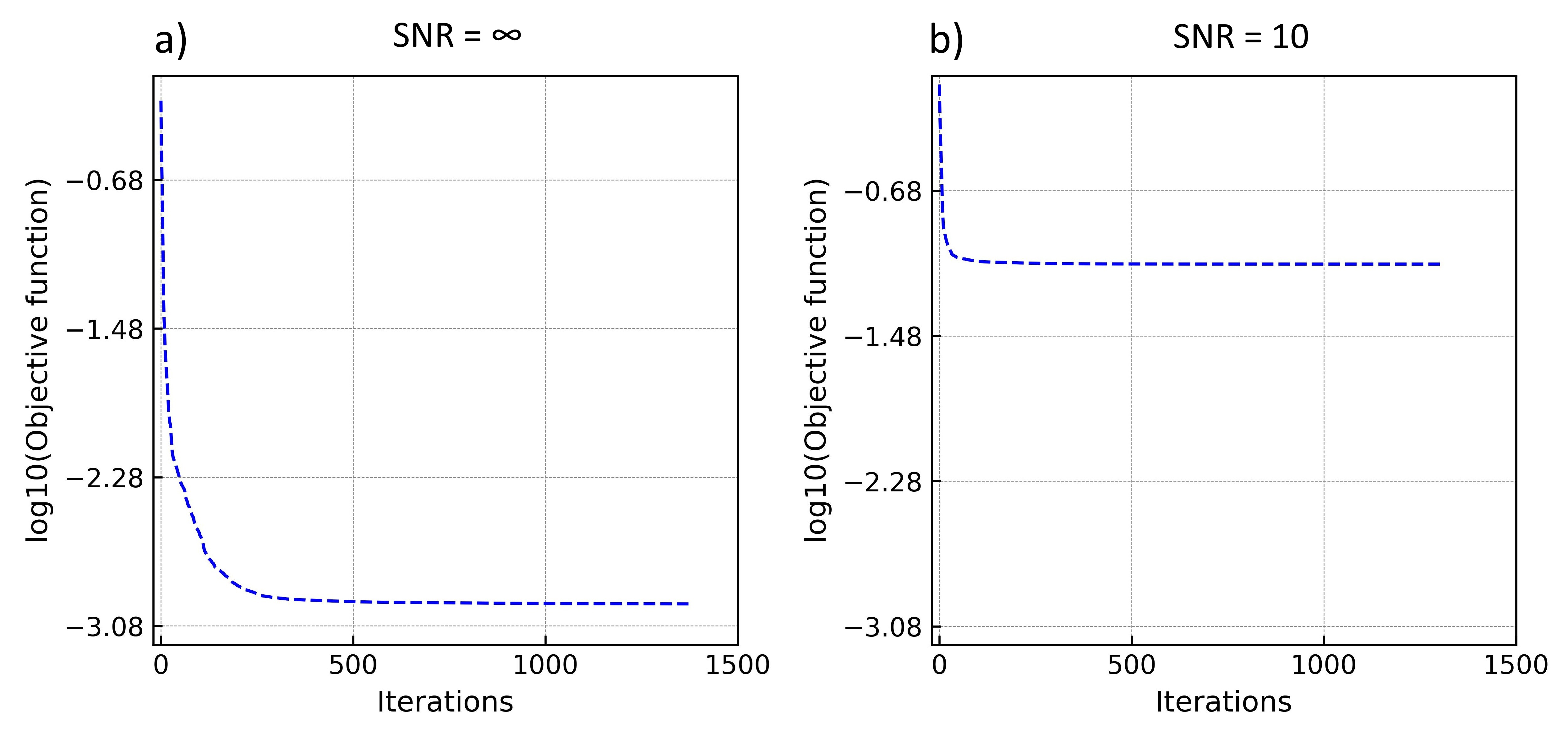}
    \caption{Convergence plot of the inversion in \new{(a) SNR = $\infty$ and (b) SNR = 10 showing $\log_{10}$} of the objective function versus iteration numbers.}
    \label{fig:5}
\end{figure}

The convergence behavior of the inversion results presented in example 2 is again assessed by plotting $J_{\mathrm{tot}}$ as a function of iteration number (Figures~\ref{fig:5}a and \ref{fig:5}b), on a logarithmic scale. The decrease in the objective function is nearly three orders of magnitude on a linear scale. However, the reduction is less pronounced in the noisy case, resulting in a higher converged objective-function value than in the noise-free case. Although the reduction in the objective function is smaller for the noisy case, the inverted elastic properties remain in good agreement with the true models.\par 

\begin{table}[ht]
\centering
\caption{\new{Example 2: Performance metrics for the inverted elastic properties under different noise levels.}}
\label{tab:2}
\begin{tabular}{llcccc}
\hline
S/N & Property & MSE & MAE & $R^2$ & $r$ \\
\hline
\multirow{3}{*}{$\infty$}
 & $V_P$  & 0.0195 & 0.1026 & 0.661 & 0.824 \\
 & $V_S$  & 0.0093 & 0.0713 & 0.682 & 0.834 \\
 & $\rho$ & 0.0012 & 0.0256 & 0.852 & 0.927 \\
\hline
\multirow{3}{*}{10}
 & $V_P$  & 0.1027 & 0.2546 & 0.475 & 0.842 \\
 & $V_S$  & 0.0290 & 0.1394 & 0.395 & 0.754 \\
 & $\rho$ & 0.0086 & 0.0702 & 0.585 & 0.768 \\
\hline
\end{tabular}
\end{table}

\begin{figure}[H]
    \centering
    \includegraphics[scale = 0.61]{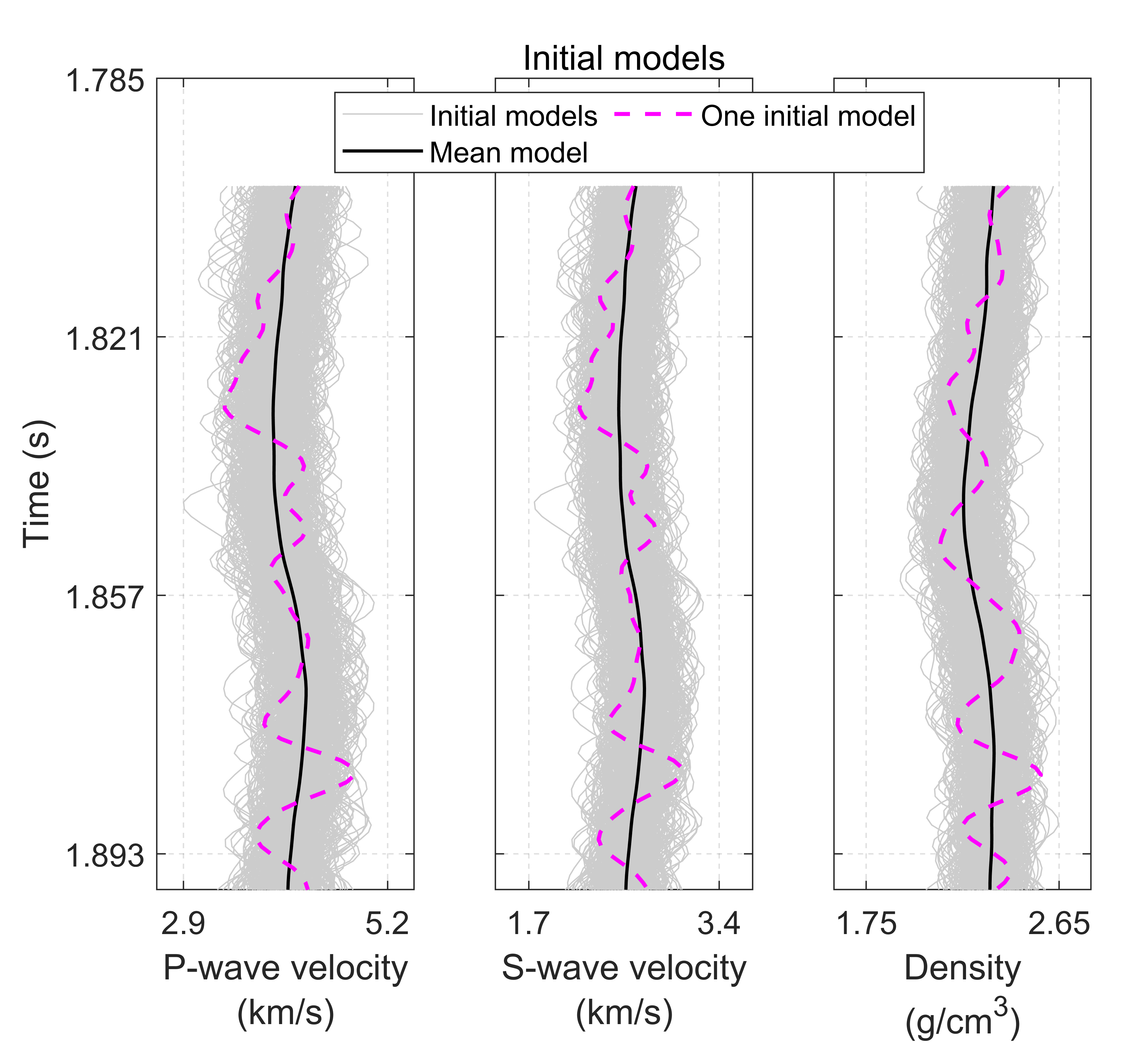}
    \caption{\new{Five hundred Monte Carlo initial models for the initial-model sensitivity test (Example~3); observed data are identical for all inversions.} The mean of all the initial realizations (in solid black), one of the initial realizations (in magenta dashed), and the initial models (in light gray) are also plotted.}
    \label{fig:initial_ensem}
\end{figure}

\begin{figure}[H]
    \centering
    \includegraphics[scale = 0.5]{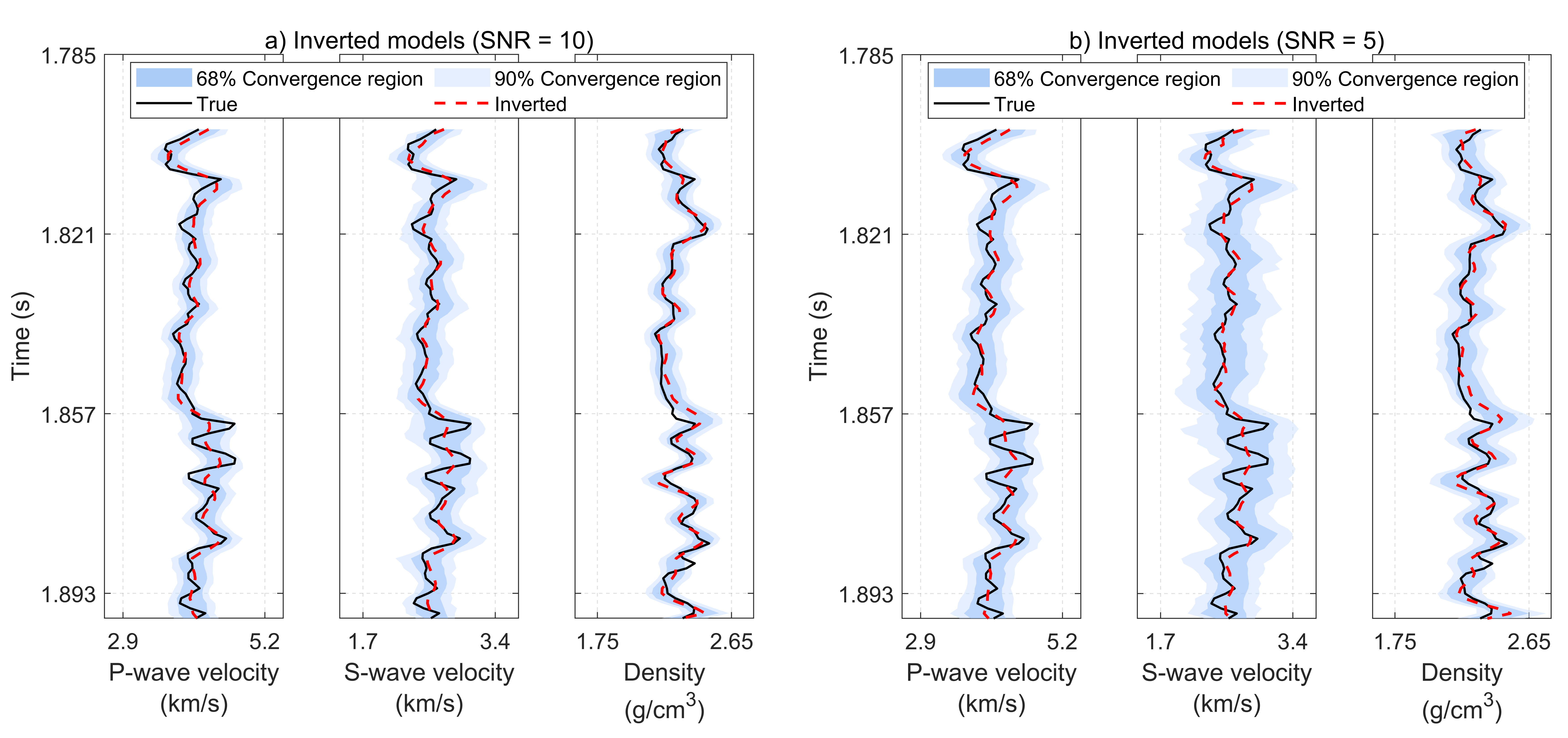}
    \caption{\new{Corresponding 500 inverted models from Example~3 (initial-model sensitivity).} The true model (in solid black), mean of the inversion models (in dashed red), and 68$\%$ and 90$\%$ convergence regions of inverted models are also plotted.}
    \label{fig:inver_ensemble}
\end{figure}

\new{The deterministic Examples~1--2 do not quantify how results depend on the starting model. We therefore perform a controlled initial-model sensitivity test: the observed seismic data (SNR~=~10 and = 5) and Zoeppritz forward operator are identical for all ensemble members, and only the starting elastic models are varied. Example~3 uses the same $J_{\mathrm{tot}}$ and parametrized $(v_p,\:\gamma,\:\rho)$ inversion as Example~2 (Section~\nameref{sec:constraints_reg}). We run the inversion 500 times with distinct initial models generated by Monte Carlo sampling, which isolates dependence on the starting model from variability due to data noise or forward-model error. A combined ensemble over both noise realizations and starting models is not performed here.} To achieve this, we used the 1D well-log dataset and the seismic model used in Figures \ref{fig:2} and \ref{fig:4}, respectively. Prior elastic models are created using Monte Carlo sampling with spatial correlation based on prior means of P- and S-wave velocities and density. The 500 resulting initial realizations, along with a single initial model (magenta dashed line) and the means of all initial models, are illustrated in Figure \ref{fig:initial_ensem}. The inversion was run 500 times, each with a fixed maximum iteration limit of 800, producing an ensemble of 500 inverted models for the P-wave velocity, the S-wave velocity, and the density. \new{Figure \ref{fig:inver_ensemble}a and b show the true models (black solid lines), the mean of all 500 inverted models (red solid lines), and the 90$\%$ and 68$\%$ convergence regions (light and dark blue shaded regions) of the P-wave velocity, S-wave velocity, and density for noise levels corresponding to $\mathrm{SNR}=10$ and $\mathrm{SNR}=5$, respectively. Despite the increased noise level, the convergence regions are centered around the true models, and the ensemble means also demonstrate strong consistency with the true values. Furthermore, although the convergence regions are wider for $\mathrm{SNR}=5$ than for $\mathrm{SNR}=10$, particularly for the S-wave velocity model, the median inverted models remain in good agreement with the true elastic-property profiles.}  \par

\begin{figure}[H]
    \centering
    \includegraphics[scale = 0.27]{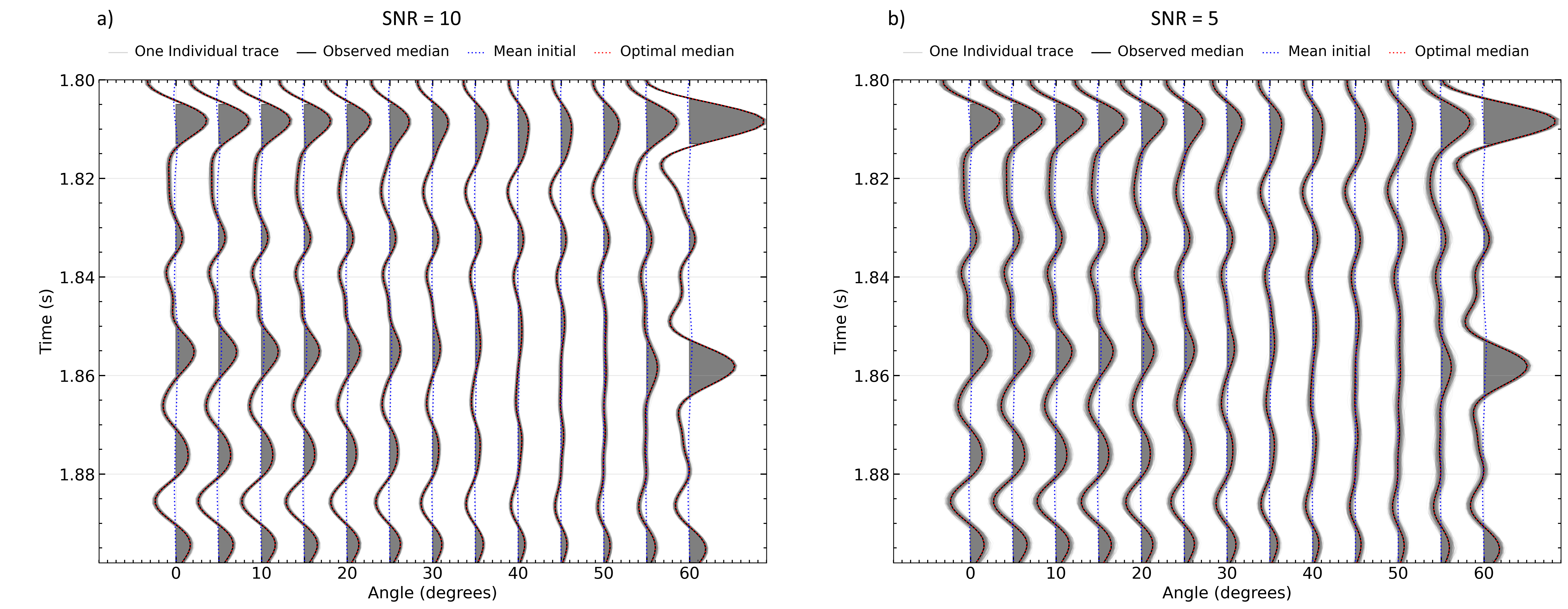}
    \caption{\new{Observed and optimal seismic traces for the ensemble-based inversion results; (a) and (b) show the results for $\mathrm{SNR}=10$ and $\mathrm{SNR}=5$, respectively. Black, red, and blue curves correspond to the median observed data, median modeled data, and modeled response from the mean initial model, respectively, while gray curves represent individual ensemble realizations.}}
    \label{fig:8}
\end{figure}

\begin{figure}[H]
    \centering
    \includegraphics[scale = 0.60]{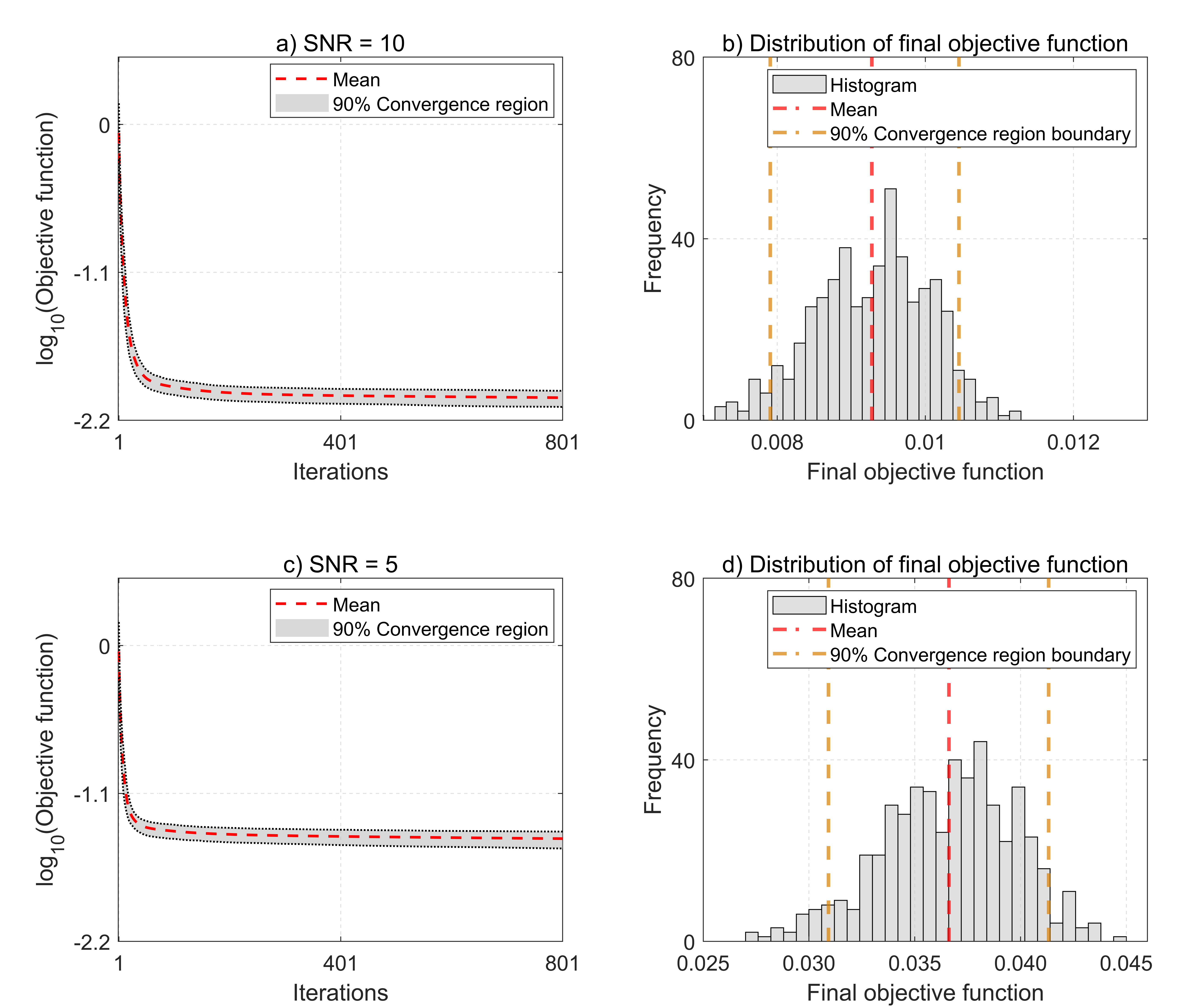}
    \caption{\new{(a) and (c) show the mean and 90\% convergence region of the $\log_{10}$ objective-function values while (b) and (d) are the distributions of the final objective-function values for $\mathrm{SNR}=10$ and 5, with the ensemble mean and 90\% bounds.}}
    \label{fig:9}
\end{figure}

\new{Figures \ref{fig:9}a and \ref{fig:9}c show that the convergence analysis reveals that the set of 500 inversion realizations consistently approaches a stable minimum of the objective function, with the mean trajectory exhibiting rapid decline and subsequent stabilization for both noise levels. Compared with the $\mathrm{SNR}=10$ case, the $\mathrm{SNR}=5$ case exhibits a smaller reduction and a broader convergence region, reflecting the increased variability among the inversion realizations under stronger noise contamination. The distribution of the final values of the objective functions (Figures \ref{fig:9}b and \ref{fig:9}d) is tightly concentrated, with the mean of the ensemble within the 90$\%$ region, which confirms that the inversion converges to a similar misfit minimum for diverse starting models, while the inverted model spread in Figure~\ref{fig:inver_ensemble} quantifies the remaining dependence on the initial model. The final objective-function distributions further show that a lower SNR leads to a wider spread and higher final misfit values.}

\begin{figure}[H]
    \centering
    \includegraphics[scale = 0.08]{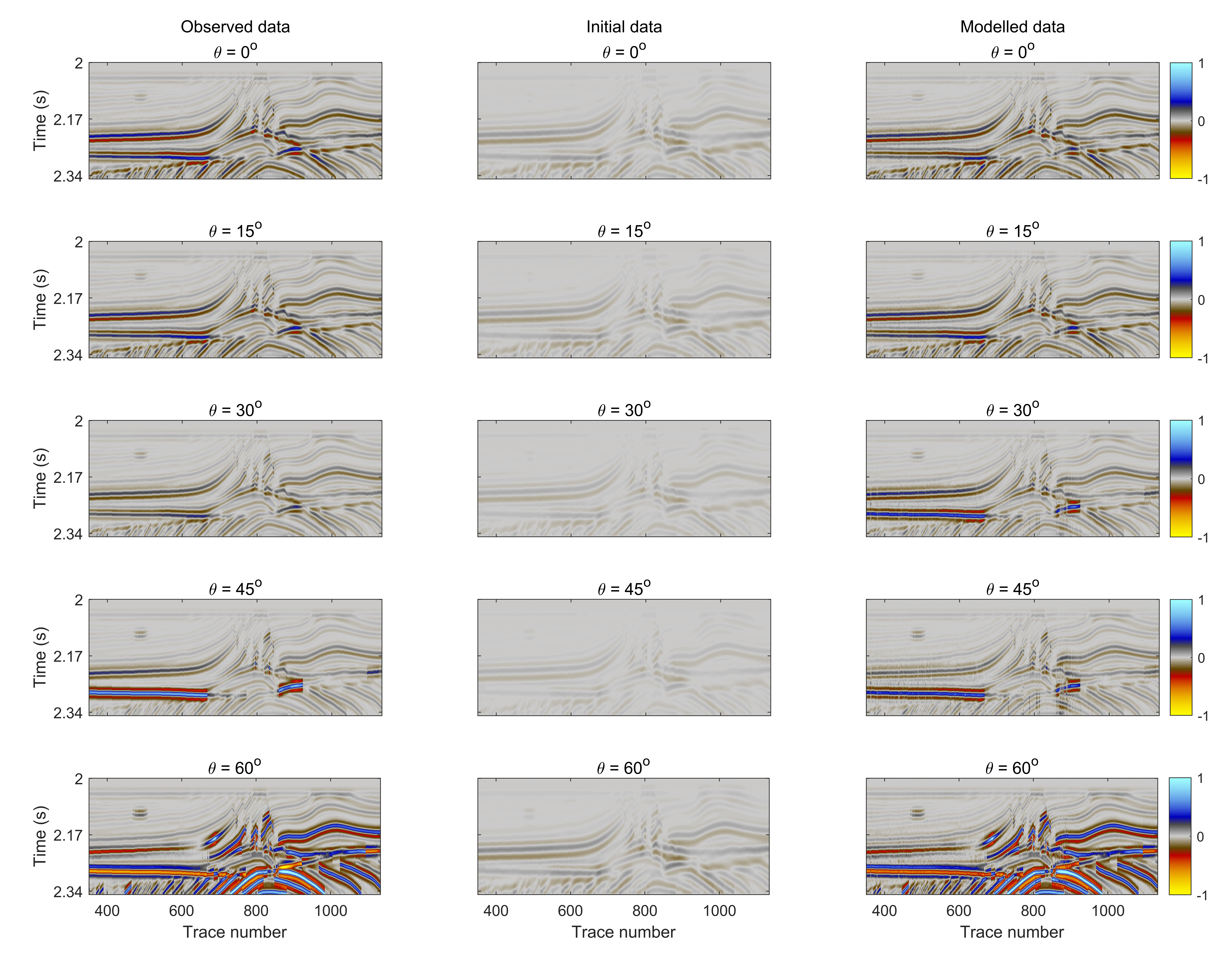}
    \caption{\new{The 2D observed seismic data, modeled seismic data obtained from the initial models, and optimal data generated from the inverted elastic-property models are represented at incidence angles of $0^\circ$, $15^\circ$, $30^\circ$, $45^\circ$, and $60^\circ$ for the Marmousi model.}}
    \label{fig:Marmousi_Seis}
\end{figure}

\new{Example~4 tests deterministic 2D recovery under structural complexity and regularization. The Marmousi model is included because its large impedance contrasts and structurally complex interfaces provide a demanding test of nonlinear Zoeppritz reflectivity at wide incidence angles (0$\degree$--60$\degree$ with the interval of 5$\degree$), including post-critical regimes where linearized approximations degrade. As in Examples~1--3, a single noisy data realization and a GMM-based initial model are used; this example therefore assesses recovery quality and lateral continuity rather than ensemble sensitivity. Synthetic seismic data are generated using the exact Zoeppritz equations, shown in Figure \ref{fig:Marmousi_Seis} with an interval of 15$\degree$. A 45 Hz zero-phase Ricker wavelet is used to band-limit the data, and multiplicative random noise was added to the synthetic seismic data. Each seismic sample was perturbed by a uniformly distributed random factor, resulting in amplitude variations of up to ±5$\%$ relative to the noise-free seismic data. This procedure generated noise-contaminated datasets that were subsequently used in the inversion experiments.} The 2D initial elastic property models are generated using a Gaussian mixture distribution \citep{grana2021seismic}. \new{Regularization follows Section~\nameref{sec:constraints_reg} with the $\gamma=v_s/v_p$ parameterization.} 

\begin{figure}[H]
    \centering
    \includegraphics[scale = 0.08]{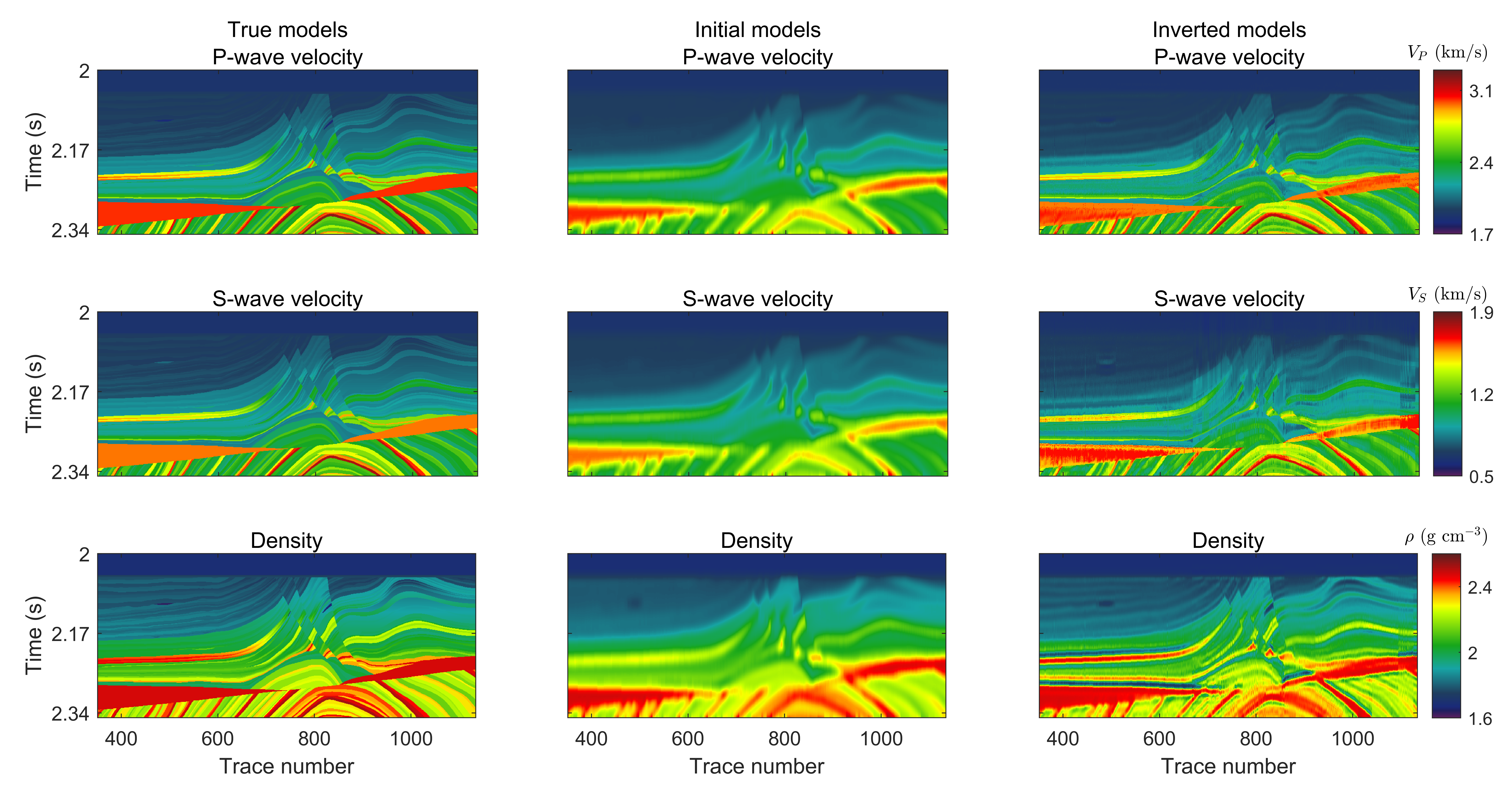}
    \caption{The 2D \new{true, initial and inverted} models of P-wave velocity, S-wave velocity, and density.}
    \label{fig:Marmousi_inver}
\end{figure}

The seismic inversion is run trace-by-trace, and the true, initial, and inverted models of P-wave velocity, S-wave velocity, and density are shown in Figure \ref{fig:Marmousi_inver}. Figure \ref{fig:Marmousi_inver} shows that the inversion method accurately predicts the model properties, especially in cases where the subsurface contains abrupt velocity changes and complex structural geometry. The comparison of the true and inverted models show that the results demonstrate a close match to the true model and exhibit coherent lateral trends, validating the geological continuity in the data. Despite the nonlinearity of the forward model, the estimated elastic parameters converge toward the true subsurface properties, highlighting the effectiveness of the optimization strategy.

\subsection{Field data examples}

\new{The field example assesses applicability to real data with a single well-log-based initial model and does not repeat the initial-model ensemble of Example~3. Field partial-angle stacks are treated within the same convolutional framework as the synthetic tests, i.e., as primary PP angle gathers after conventional prestack processing, without explicit two-dimensional wave-equation modeling in the inversion loop.}
The seismic amplitude-versus-offset inversion method is applied to seismic field data from the Troll field in the Norwegian North Sea. The Troll field is one of the largest offshore gas and oil fields in the North Sea, located west of Norway on the Horda Platform, which contains massive gas accumulations with a thin oil rim. A vertical slice from inline 900 is extracted for testing the proposed inversion, as it exhibits pronounced lateral and vertical structural variations. \new{The seismic data comprises three-angle stacks (near, mid, and far) with a sampling interval of 4 milliseconds. The near-, mid-, and far-angle stacks correspond to central incidence angles of approximately $8^\circ$, $22^\circ$, and $37^\circ$, respectively.} The wavelets were extracted statistically by computing and averaging the amplitude spectra of all traces, smoothing the composite spectrum, and converting it to the time domain to obtain compact zero-phase wavelets. Low-frequency models of the well log data are used as initial models, and seismic inversion is performed trace-by-trace.  \par

\begin{figure}[H]
    \centering
    \includegraphics[scale = 0.5]{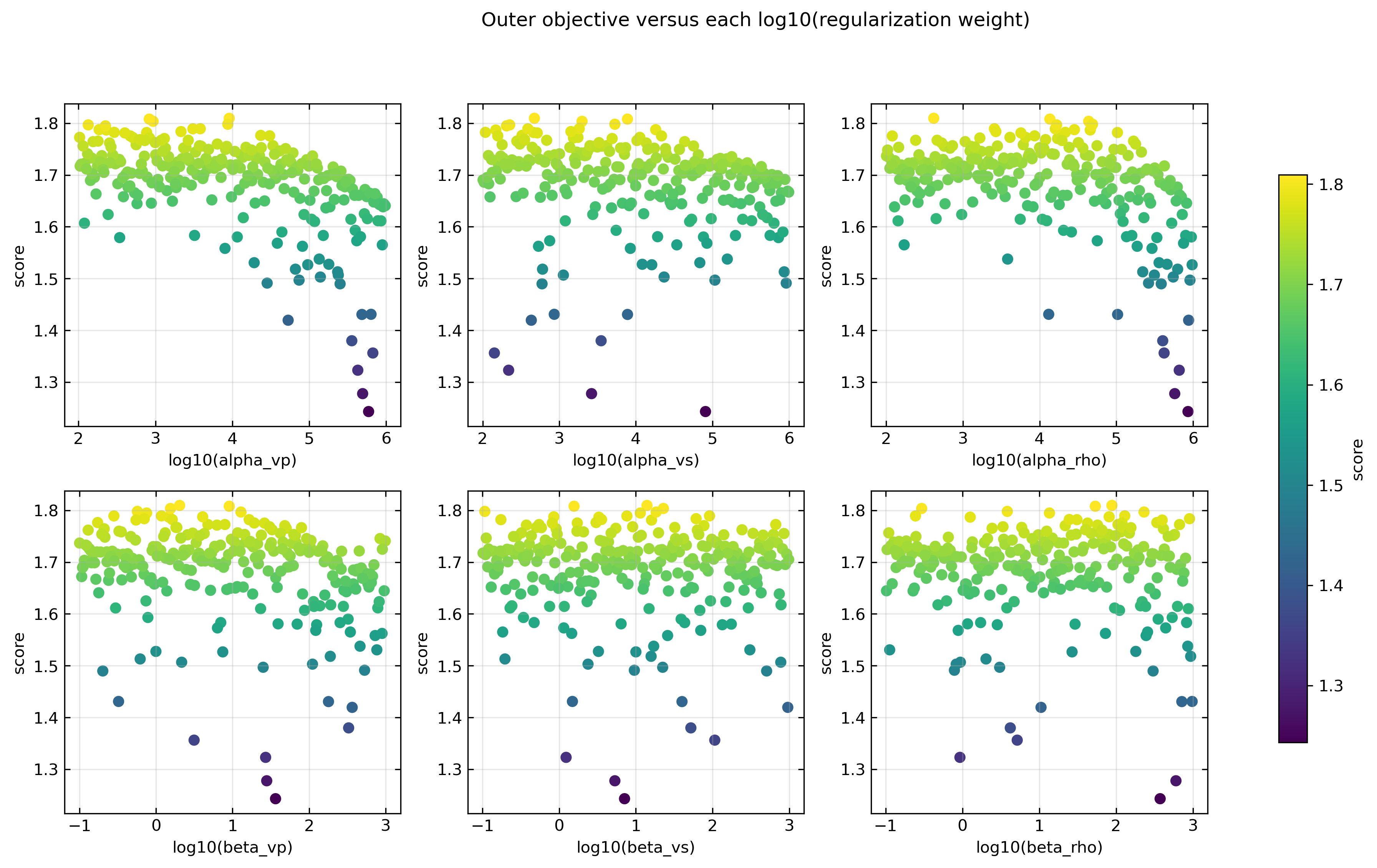}
    \caption{\new{Outer-loop optimization of the six regularization weights. The objective score is plotted against the logarithm of the TV weights (top row) and Tikhonov weights (bottom row).}}
    \label{fig:regul}
\end{figure}

\new{Both the Tikhonov and total-variation (TV) constraints are applied to stabilize the inversion and obtain geologically plausible elastic property models. The quality of inversion results depends on the choice of regularization weights; an outer-loop parameter search was performed to identify the optimal constraints. For each candidate set of weights, the inversion was executed and evaluated using a composite score that combines the correlation between the inverted and well-log elastic properties with the seismic data fit. Figure~\ref{fig:regul} shows the resulting objective score as a function of the six regularization weights. The upper panels correspond to the TV weights ($\alpha_{vp}$, $\alpha_{vs}$, and $\alpha_{\rho}$), whereas the lower panels correspond to the Tikhonov weights ($\beta_{vp}$, $\beta_{vs}$, and $\beta_{\rho}$). Each point describes a separate inversion run, and warmer colors define higher scores. The optimal parameter set was identified as $\alpha_{vp}=9.11\times10^{3}$, $\alpha_{vs}=4.73\times10^{2}$, $\alpha_{\rho}=4.17\times10^{2}$, $\beta_{vp}=2.04$, $\beta_{vs}=13.96$, and $\beta_{\rho}=88.24$, yielding a score of 1.81, a mean correlation coefficient of 0.88, and a data-fit value of 0.93. These weights were subsequently used in the final inversion.}

\begin{figure}[H]
    \centering
    \includegraphics[scale = 0.5]{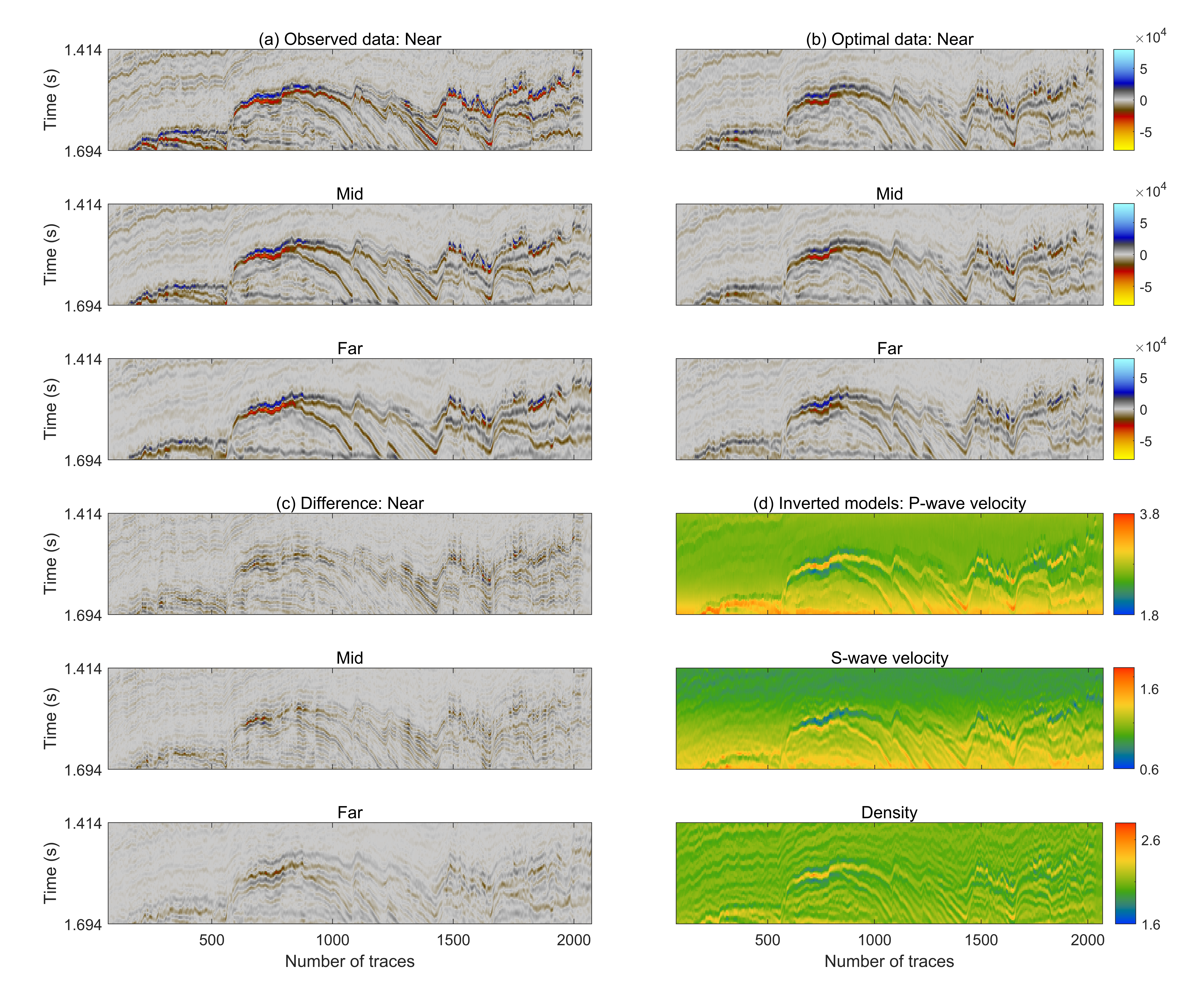}
    \caption{\new{Troll Field data and inversion results. Panel (a) presents the observed near-, mid-, and far-angle seismic stacks. Panel (b) displays the corresponding optimal seismic data generated from the inverted elastic-property models. Panel (c) shows the residuals (observed minus optimal data) for the near-, mid-, and far-angle stacks. Panel (d) shows the inverted elastic property models.}}
    \label{fig:2DTroll}
\end{figure}

\new{Figure~\ref{fig:2DTroll} shows the observed seismic data, optimal responses, residuals, and inverted elastic properties. Panel (a) presents the observed near-, mid-, and far-angle seismic stacks, while panel (b) displays the corresponding modeled data generated from the inverted elastic properties. Panel (c) presents the residuals (observed minus modeled data) for the near-, mid-, and far-angle stacks. Panel (d) shows the inverted elastic property models, including the P wave velocity, the S wave velocity, and the density.} The inversion results yield high-resolution elastic models that are consistent with the seismic data and remain constrained within realistic physical limits. \new{The optimal seismic data closely agree with the observed near-, mid-, and far-angle stacks. The residuals are generally small and do not exhibit systematic patterns, indicating that the inverted elastic properties provide an adequate fit to the observed seismic data.} The results indicate that the inversion recovers elastic-property variations that are consistent with the observed angle stacks and with the available well-log constraints. Major reflectors are reproduced clearly, whereas interpretation of smaller-scale features should be made with caution because independent validation is limited.\par 

\begin{figure}[H]
    \centering
    \includegraphics[scale = 0.5]{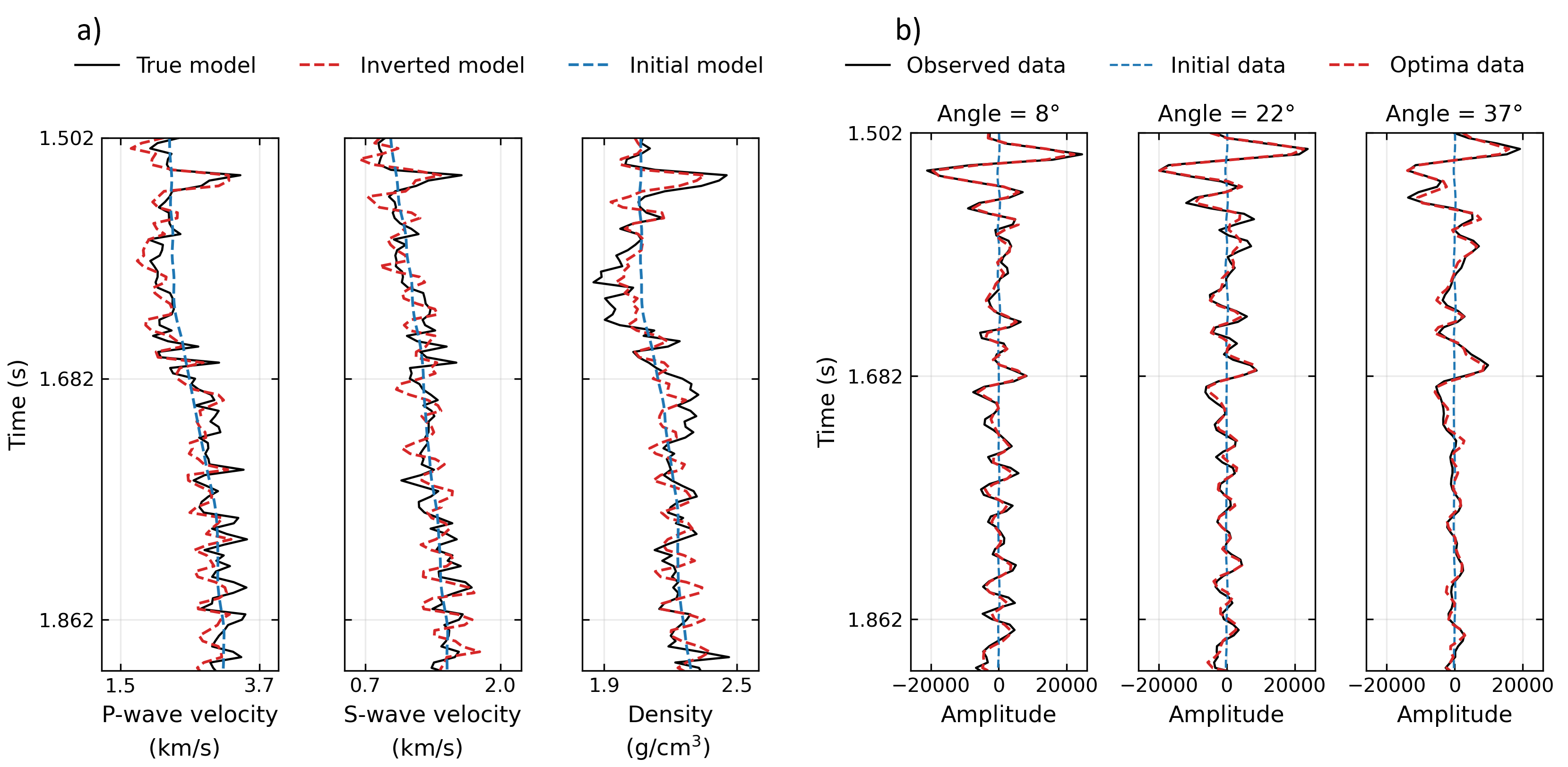}
    \caption{\new{(a) Comparison between actual, initial, and predicted elastic property models at the well location and (b) corresponding observed-to-optimal seismic trace fit.}}
    \label{fig:1DTroll}
\end{figure}

\new{To further assess the inversion performance, the results were shown at the well location (trace 937). Figure~\ref{fig:1DTroll}(a) compares the inverted P-wave velocity, S-wave velocity, and density models with the corresponding true logs at the well location (31/2-1). Figure~\ref{fig:1DTroll}(b) shows the associated seismic data fit, comparing the observed and modeled near-, mid-, and far-angle traces extracted at the same location. The predicted elastic models are in close agreement with the true well logs, while the optimal seismic traces closely match the observed data across all angle stacks, showing a reliable property prediction.}

\section{Discussion}

\new{The results demonstrate that combining the exact Zoeppritz equations with a discretized adjoint-state formulation provides an effective framework for nonlinear prestack AVO inversion. The synthetic and field-data examples show that the proposed approach accurately recovers elastic parameters while preserving the full nonlinear reflectivity behavior. In contrast to conventional linearized AVO approximations, the exact formulation remains applicable in the presence of strong elastic contrasts, large incidence angles, and post-critical-angle reflections. The following sections discuss the impact of the forward-model approximation, angular sampling, efficient gradient computation strategy, and the limitations of the proposed framework.}

\subsection{Zoeppritz versus linearized reflectivity}\label{sec:zoep_aki}

\begin{figure}[H]
    \centering
    \includegraphics[width=0.85\textwidth]{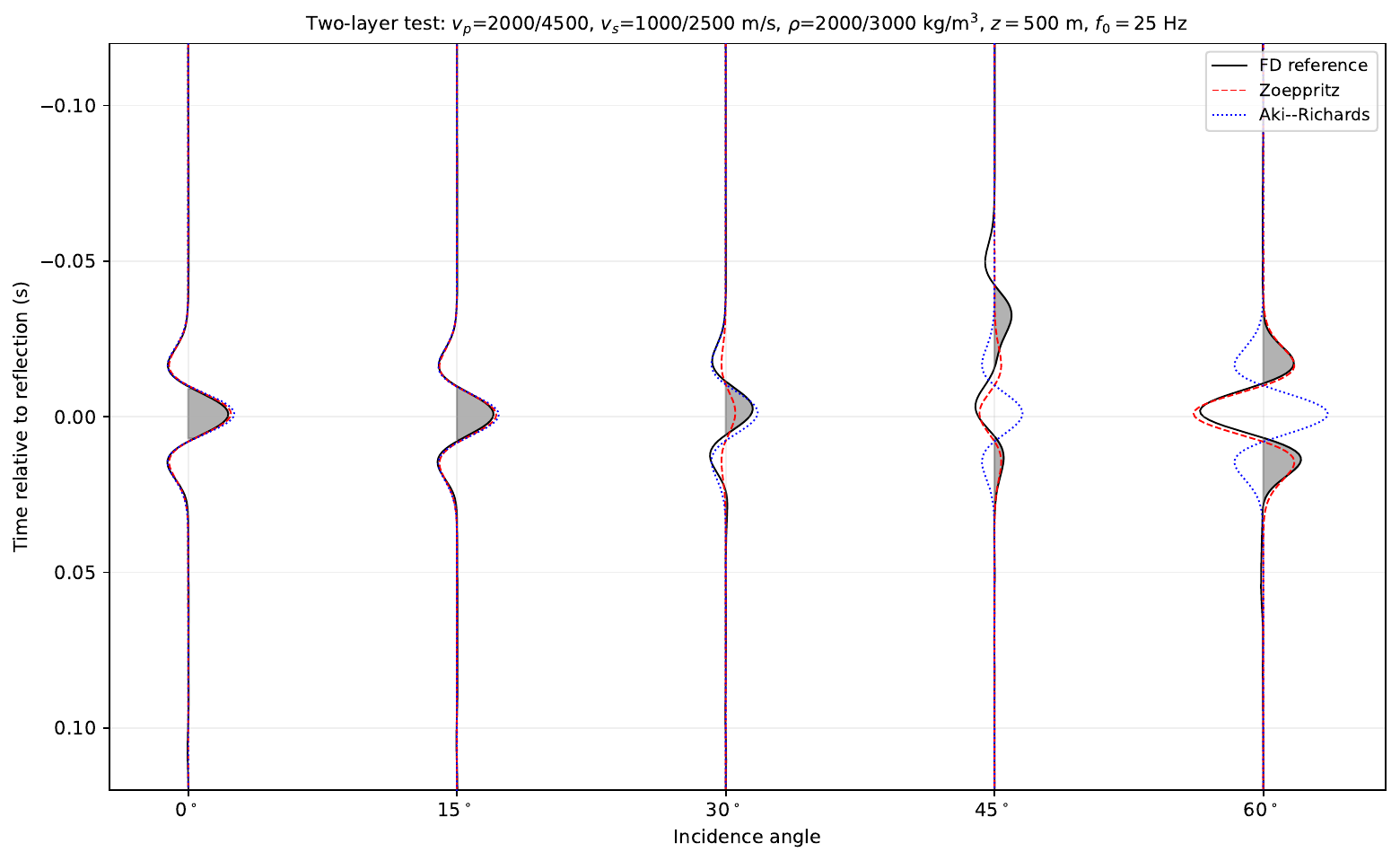}
    \caption{\new{Wiggle comparison of reference wave-equation partial stacks (black) with convolutional synthetics from exact Zoeppritz (red dashed) and linearized Aki--Richards (blue dotted) on a high-contrast two-layer model ($v_{p1},v_{s1},\rho_1)=(2000,1000,2000)$, $(v_{p2},v_{s2},\rho_2)=(4500,2500,3000)$, reflector at 500~m). All five incidence angles $0\degree$--$60\degree$ are shown in one panel with three overlapping wiggles per angle; time windows are centered on the reflection and span $\pm 3/f_0$~s ($f_0=25$~Hz). Reference data are from two-dimensional elastic finite-difference modelling (}\citep{virieux1986}\new{) with line-source to point-source geometrical-spreading correction; synthetics follow equation~}\eqref{eqn:1}\new{ with $\operatorname{Re}(R_{PP})$ for the real wavelet. At $30\degree$ and $45\degree$, head-wave energy interferes with the reflection on the reference traces.}}
    \label{fig:zoep_aki}
\end{figure}

\new{To quantify the benefit of the exact forward model over linearized AVO, we compare convolutional synthetic angle gathers from Zoeppritz and from the linearized Aki--Richards approximation} \citep{aki2002quantitative} \new{with reference partial-angle stacks from two-dimensional elastic finite-difference modelling. Reference wavefields use the staggered-grid velocity--stress scheme of} \citep{virieux1986} \new{on a two-layer elastic model with a horizontal interface at $z=500$~m and properties $(v_{p1},v_{s1},\rho_1)=(2000,1000,2000)$ and $(v_{p2},v_{s2},\rho_2)=(4500,2500,3000)$~m/s and kg/m$^3$ (SI units) respectively. The scattered response is obtained by subtracting a homogeneous-background simulation and muting samples below the reflection arrival. Amplitudes are corrected from a two-dimensional line-source to a three-dimensional point-source geometrical spreading so that the reference gathers follow point-source decay, and an analytical spherical divergence factor is applied at the reflector before comparison. A 25~Hz zero-phase Ricker wavelet is used for both the reference experiment and the convolutional synthetics in equation~}\eqref{eqn:1}\new{;~Zoeppritz synthetics use complex post-critical $R_{PP}$, whereas Aki--Richards uses the linearized approximation. Synthetics are time-shifted to the reflection traveltime at each incidence angle before comparison. Figure~}\ref{fig:zoep_aki}\new{~shows a single wiggle plot with incidence angles $\theta = 0\degree,\:15\degree,\:30\degree,\:45\degree,\:60\degree$ (three overlapping wiggles per angle: reference, Zoeppritz, Aki--Richards). At $\theta = 30\degree$ and $45\degree$, the reference finite-difference traces exhibit interference between the reflected arrival and the head wave; this contribution is absent from the convolutional synthetics, which model only the interface reflectivity in equation~}\eqref{eqn:1}\new{. At $\theta\ge 45\degree$, the interface is post-critical for the upper medium, where Zoeppritz remains defined in the complex domain but the linearized approximation degrades. In Figure \ref{fig:zoep_aki}, we focus on forward-model fidelity rather than paired inversions with the Aki--Richards adjoint.}

\begin{figure}[H]
    \centering
    \includegraphics[scale = 0.5]{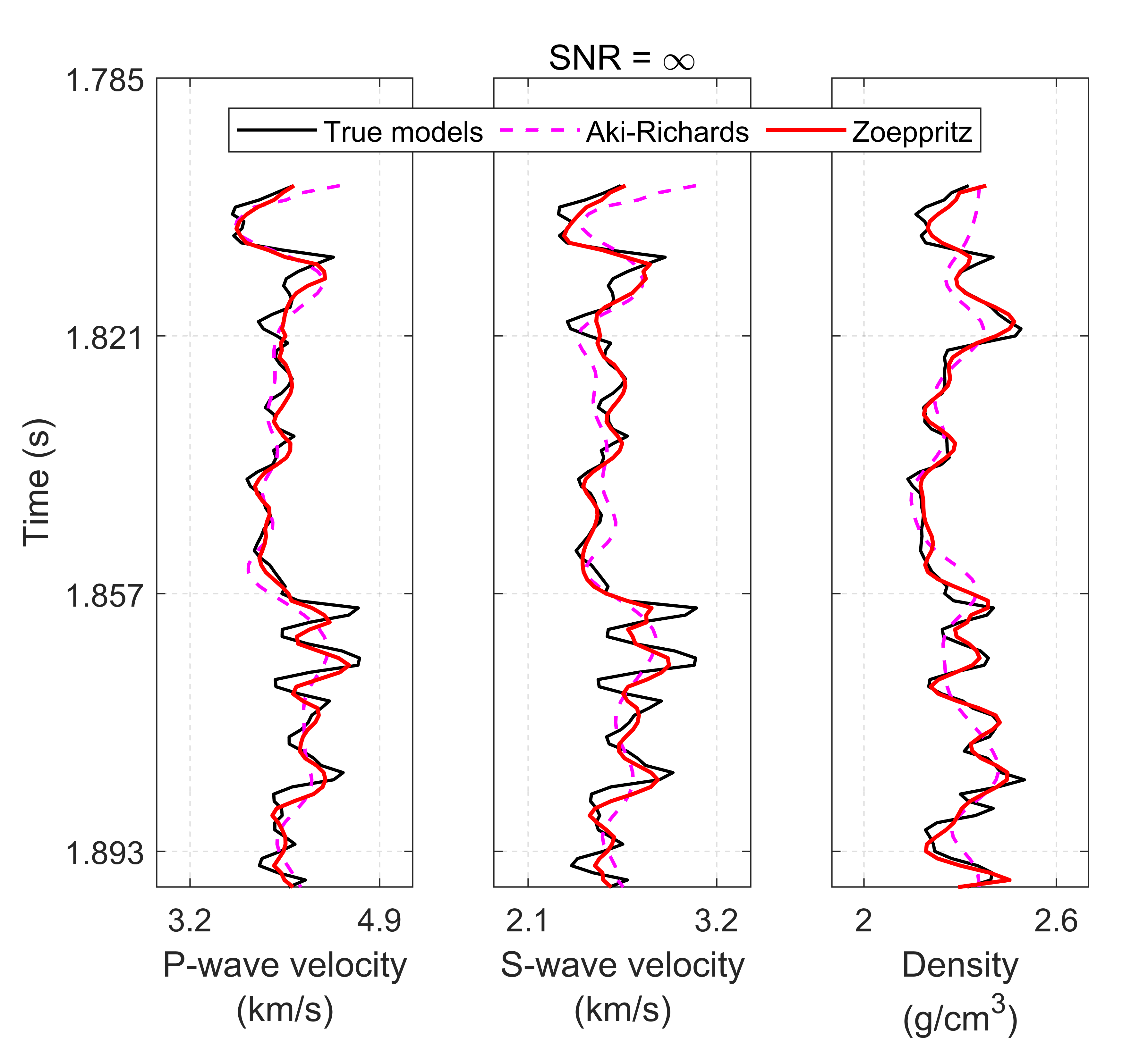}
    \caption{\new{Comparison of inversion results using the exact Zoeppritz equations and the linearized Aki--Richards approximation obtained using identical inversion settings, showing improved elastic-property recovery with the exact Zoeppritz formulation.}}
    \label{fig:16}
\end{figure}

\new{To assess the practical implications of the forward-model approximation, we further compare inversion results obtained using the linearized Aki--Richards forward model of} \citet{ahmed2022constrained} \new{and the proposed exact Zoeppritz formulation. For a fair comparison, both methods employ the same inversion framework, adjoint-state optimization strategy, regularization, initial model, and noise-free synthetic data (SNR = $\infty$) used in example 1; the only difference is the forward operator, which is based on the equivalent formulation derived in this work. As shown in Figure \ref{fig:16}, the inversion based on the exact Zoeppritz equations recovers the elastic parameters more accurately, particularly in regions of strong impedance contrast where the linearized approximation introduces modeling errors. This improvement is consistent with the closer agreement between the exact Zoeppritz response and the finite-difference reference, demonstrating that the enhanced forward-model accuracy directly translates into improved inversion performance.}

\subsection{Computational aspects of gradient evaluation}

\new{Gradient-based nonlinear inversion can be implemented using finite-difference perturbations, explicit Jacobian (or Fréchet-derivative) formulations, or adjoint-state methods. Finite-difference gradients are conceptually straightforward but require repeated forward-model evaluations for each perturbed model parameter, causing the computational cost to increase linearly with the number of unknowns. Alternatively, several recent nonlinear AVO inversion methods based on the exact Zoeppritz equations derive analytical Jacobian, Hessian, or Fréchet-derivative expressions to compute parameter sensitivities} \citep{liu2019accurate, chai2020elastic, Bao2021, xu2023prestack}\new{. In these approaches, the gradient is obtained by evaluating the sensitivity of the forward operator with respect to each model parameter at every model sample and incidence angle, assembling the corresponding sensitivity matrices, and subsequently multiplying these by the data residual. Although these formulations provide exact analytical derivatives, the explicit construction and storage of sensitivity operators become increasingly demanding as the number of model parameters, incidence angles, and model samples increases.} \par

\begin{figure}[H]
    \centering
    \includegraphics[scale = 0.5]{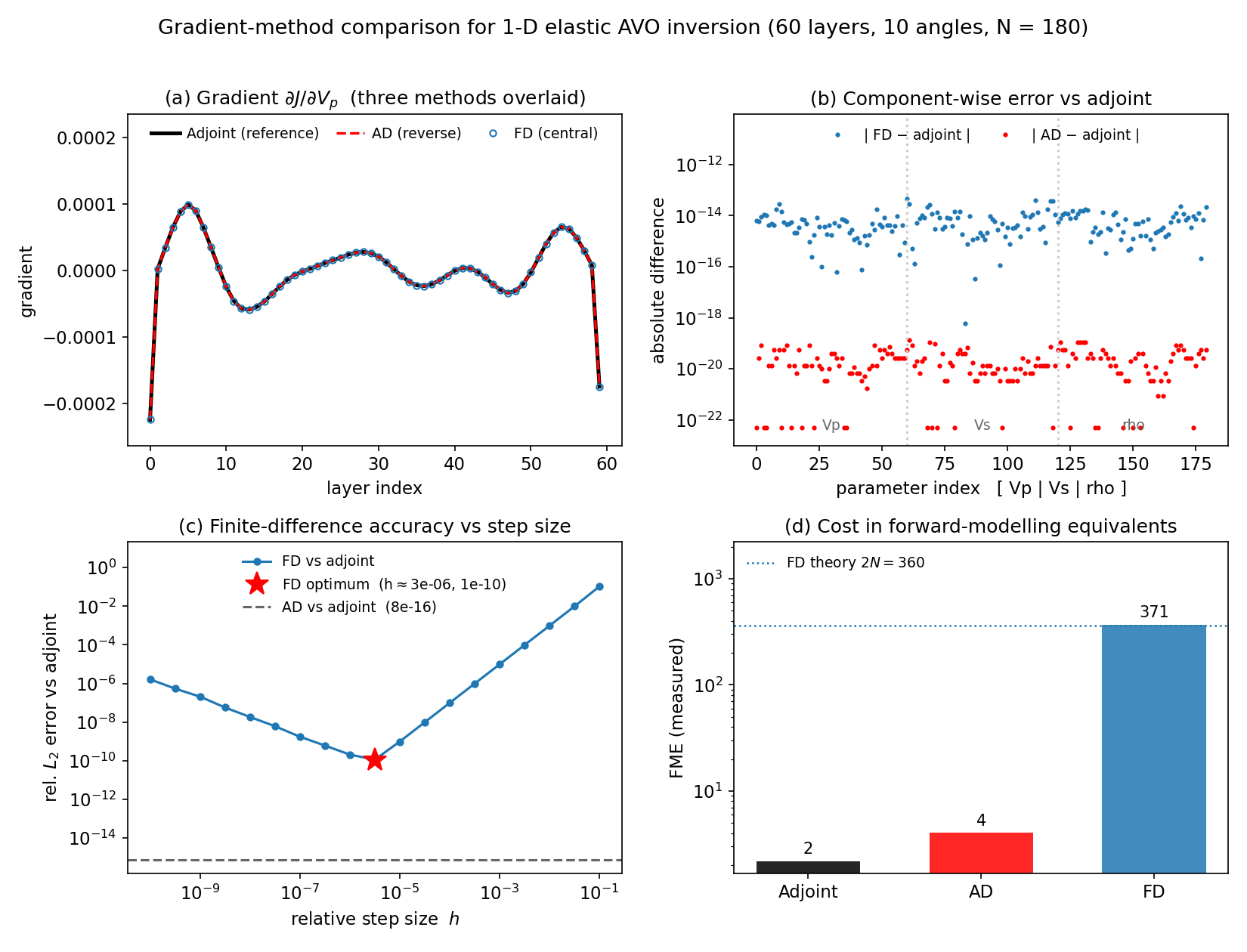}
    \caption{\new{Comparison of three methods for computing the gradient of the least-squares AVO misfit with respect to the elastic model ($V_P$, $V_S$, and $\rho$) for a 1-D layered problem. (a) Adjoint-state, reverse-mode automatic differentiation (AD), and central finite-difference (FD) gradients ($\partial J/\partial V_P$ shown). (b) Component-wise absolute differences relative to the adjoint-state gradient. (c) Relative FD error as a function of perturbation step size. (d) Computational cost is expressed in forward-model equivalents (FME), where one FME corresponds to a single forward-model evaluation.}}
    \label{fig:17}
\end{figure}

\new{The proposed adjoint-state formulation avoids repeated finite-difference (FD) perturbations, explicit Jacobian assembly, and the computational overhead associated with automatic differentiation by deriving a discrete adjoint system directly from the reformulated exact Zoeppritz forward operator. Instead of constructing and storing sensitivity matrices, the gradient is computed through a forward and an adjoint recursion, yielding matrix-free sensitivities with respect to $v_p$, $v_s$, and $\rho$. Consequently, the computational complexity grows much more favorably for large-scale multilayer inversion problems while preserving the full nonlinear Zoeppritz physics, including the treatment of complex-valued post-critical-angle reflections. To assess the accuracy and the computational efficiency of the proposed adjoint-state implementation, we compare it with the conventional finite-difference approach and the automatic differentiation (AD, reverse mode) for gradient computation. Figure~\ref{fig:17} shows the comparison of three methods for computing the gradient of the least-squares AVO misfit with respect to the elastic model ($v_p$, $v_s$, $\rho$) for a 1-D layered problem (60-layer model, 10 incidence angles spanning $0^\circ$--$45^\circ$, $N=180$ unknowns; exact Zoeppritz reflectivity convolved with the source wavelet). The analytic adjoint-state gradient is taken as the accurate reference. The three methods return the same gradient, but differ sharply in accuracy and computational cost. Automatic differentiation (AD) reproduces the analytic adjoint-state gradient to machine precision (relative $L_2$ error $8\times10^{-16}$), confirming that both evaluate the exact derivative of the forward operator. The finite-difference (FD) gradient agrees only to $2\times10^{-10}$ and, being a numerical approximation, is limited to a narrow range of perturbation step sizes, with the best attainable error of approximately $10^{-10}$ near $h=3\times10^{-6}$, bounded by truncation error for large $h$ and round-off error for small $h$. In terms of computational cost, the adjoint-state and AD methods evaluate the complete gradient using only 2 and 4 forward-model equivalents (FME), respectively, independent of the number of model parameters. In contrast, the FD method requires approximately $2N$ forward-model evaluations (371 FME in the present example), making it approximately 171 times more expensive. The adjoint-state method, therefore, provides the preferred implementation for large-scale nonlinear Zoeppritz inversion, combining exact gradient accuracy with substantially lower computational cost. The quantitative comparison is summarized in Table~\ref{tab:3}, confirming the high accuracy and computational efficiency of the proposed adjoint-state formulation. The adjoint-state method achieves substantially lower computational cost while maintaining the same gradient accuracy, making it more suitable for large-scale nonlinear Zoeppritz inversion.} 

\begin{table}[ht]
\centering
\caption{\new{Computational cost (in forward-modelling equivalents, both the theoretical scaling and the measured value) and gradient accuracy (relative $L_2$ error against the analytic adjoint-state gradient) of the three methods for the
N = 180 parameter problem.}}
\label{tab:3}
\begin{tabular}{lccc}
\hline
Method & Relative $L_2$ error & Cost (theoretical) & Cost (measured) \\
\hline
Adjoint-state & Reference & 2 FME & 2 FME \\
Automatic differentiation & $8\times10^{-16}$ & 4 FME & 4 FME \\
Finite differences & $2\times10^{-10}$ & $2N=360$ FME & 371 FME \\
\hline
\end{tabular}
\end{table}

\subsection{Limitations and future work}

\new{The proposed inversion framework is formulated within the conventional convolutional AVO approximation and assumes that the input data are dominated by primary PP reflections obtained after standard preprocessing steps such as amplitude balancing, geometrical-spreading correction, and attenuation compensation. Consequently, multidimensional wave-propagation effects, interbed multiples, converted-wave interactions, and head-wave contributions are not explicitly modeled. Although these assumptions are common in prestack AVO inversion, they may become restrictive in complex geological environments where wave-propagation effects play an important role. Future work could therefore extend the present adjoint-state formulation to wave-equation-based forward operators while retaining the exact-Zoeppritz treatment of large-angle and post-critical-angle reflections.}

\section{Conclusions}

In this work, we present a gradient-based amplitude-variation-with-offset inversion method to estimate elastic properties using the exact Zoeppritz equation. We derive an analytical gradient of the data misfit $J_{\mathrm{data}}$ with respect to the elastic properties by applying the adjoint-state method to the exact Zoeppritz equations. The nonlinear inverse problem minimizes $J_{\mathrm{tot}}=J_{\mathrm{data}}+J_{\mathrm{Tikh}}+J_{\mathrm{TV}}$ (Section~\nameref{sec:constraints_reg}) using limited-memory quasi-Newton optimization, with logistic bounds on physical parameters and optional $V_S/V_P$ reparameterization.

\new{We applied the introduced technique to 1D and 2D synthetic datasets using different signal-to-noise ratios (SNR~=~$\infty$, 10 and 5 in the 1D tests, and random amplitude perturbations of up to ±5$\%$ relative to the noise-free seismic data were applied in the Marmousi example). Initial-model sensitivity was assessed in Example~3 using a 500-member ensemble with identical noisy data. On a high-contrast two-layer test, convolutional Zoeppritz synthetics track two-dimensional finite-difference reference gathers (with line-source to point-source spreading correction) more closely than the linearized Aki--Richards approximation at wide angles, including post-critical incidence (Figure~}\ref{fig:zoep_aki}\new{). The results demonstrate reliable recovery in the deterministic tests and convergence of diverse starting models toward a common solution in the ensemble experiment, with posterior intervals narrower than the prior but not fully calibrated as 95$\%$ uncertainty bounds.} The inversion method was then applied to seismic data from the tectonically complex part of the Troll field, the Norwegian North Sea, to evaluate its robustness and adaptability in more challenging geological settings.

\new{Appendix added: derivation of an equivalent Zoeppritz formulation used in implementation (following section).}

\bibliographystyle{apalike}
\bibliography{Bibliography.bib}
\newpage


\segappend{Derivation of an equivalent Zoeppritz formulation for forward modelling}{app:equivalent_zoep}

\new{This appendix derives an algebraically equivalent reformulation of the PP Knott-Zoeppritz equations used in the forward and adjoint computations, following the contrast-based parameterization of} \citet{lavaud1999pushing} \new{and } \citet{chavent2010nonlinear}\new{.}

\paragraph{Original Knott-Zoeppritz equations}
\new{For a P-wave incident at angle $\theta_1$ on an interface between
$(v_{p,1},v_{s,1},\rho_1)$ and $(v_{p,2},v_{s,2},\rho_2)$, the ray parameter relation is}
\begin{equation}
 \frac{\sin\theta_1}{v_{p,1}} =
 \frac{\sin\theta_2}{v_{p,2}} =
 \frac{\sin\phi_1}{v_{s,1}} =
 \frac{\sin\phi_2}{v_{s,2}} = p.
\end{equation}
\new{The Knott-Zoeppritz reflection/transmission system can be written in the form} \citep{Knott1899,Zoeppritz1919,AkiRich2010}
\begin{equation}\label{eq:app_Rpp_eps}
 R_{PP} =
 \left[\mathcal{E}F + \mathcal{G}H\,p^2\right]_{\varepsilon=-1}
 \Big/
 \left[\mathcal{E}F + \mathcal{G}H\,p^2\right]_{\varepsilon=+1},
\end{equation}
\begin{equation}
 R_{PS} =
 -2\frac{\cos\theta_1}{v_{p,1}}
 \left(
 ab + cd\,\frac{\cos\theta_2}{v_{p,2}}\frac{\cos\phi_2}{v_{s,2}}
 \right)\,
 \frac{p\,v_{p,1}}{v_{s,1}\left[\mathcal{E}F + \mathcal{G}H\,p^2\right]_{\varepsilon=+1}},
\end{equation}
\new{with}
\begin{equation}
 \mathcal{E} = b\frac{\cos\theta_1}{v_{p,1}} + \varepsilon c\frac{\cos\theta_2}{v_{p,2}},
 \quad
 F = b\frac{\cos\phi_1}{v_{s,1}} + c\frac{\cos\phi_2}{v_{s,2}},
\end{equation}
\begin{equation}
 \mathcal{G} = a - \varepsilon d\frac{\cos\theta_1}{v_{p,1}}\frac{\cos\phi_2}{v_{s,2}},
 \quad
 H = a - d\frac{\cos\theta_2}{v_{p,2}}\frac{\cos\phi_1}{v_{s,1}},
\end{equation}
\new{and}
\begin{equation}
 \begin{aligned}
 a &= \rho_2(1-2v_{s,2}^{2}p^{2}) - \rho_1(1-2v_{s,1}^{2}p^{2}), \\
 b &= \rho_2(1-2v_{s,2}^{2}p^{2}) + 2\rho_1 v_{s,1}^{2}p^{2}, \\
 c &= \rho_1(1-2v_{s,1}^{2}p^{2}) + 2\rho_2 v_{s,2}^{2}p^{2}, \\
 d &= 2(\rho_2 v_{s,2}^{2} - \rho_1 v_{s,1}^{2}).
 \end{aligned}
\end{equation}
\new{These equations are intricate and difficult to differentiate directly, which motivates the equivalent reformulation derived below.}

\paragraph{Revisiting the Knott-Zoeppritz equations}
\new{Following }\citet{lavaud1999pushing} \new{and }\citet{chavent2010nonlinear}\new{,} \new{we introduce contrast parameters}
\begin{align}\label{eq:app_contrasts}
 e_r &= \frac{\rho_2-\rho_1}{\rho_2+\rho_1}, &
 e_p &= \frac{v_{p,2}^{2}-v_{p,1}^{2}}{v_{p,2}^{2}+v_{p,1}^{2}}, &
 e_s &= \frac{v_{s,2}^{2}-v_{s,1}^{2}}{v_{s,2}^{2}+v_{s,1}^{2}},
\end{align}
\new{and background parameters}
\begin{equation}\label{eq:app_background}
 \bar\rho = \frac{\rho_1+\rho_2}{2}, \qquad
 \frac{1}{\bar v_p^{2}} = \frac{1}{2}\left(\frac{1}{v_{p,1}^{2}}+\frac{1}{v_{p,2}^{2}}\right), \qquad
 \bar v_s^{2} = \frac{v_{s,1}^{2}+v_{s,2}^{2}}{2}.
\end{equation}
\new{The geometric mean for $\bar v_p$ and the arithmetic mean for $\bar v_s$ are chosen so that the reformulated system depends on the background velocity ratio $\chi=\bar v_s/\bar v_p$ only (not on $\bar v_p$ and $\bar v_s$ separately). With $\mathcal{E}$ and $\mathcal{G}$ as above and $D_\varepsilon=\mathcal{E}F+\mathcal{G}H\,p^2$, we express the layer velocities in terms of $(e_p,e_s,\chi)$, where}
\begin{equation}\label{eq:app_chi_chavent}
 \chi = \frac{\bar v_s}{\bar v_p},
\end{equation}
\new{and define}
\begin{align}\label{eq:app_ST}
 S_1 &= \frac{2\bar v_s^{2}}{v_{p,1}^{2}} = 2\chi^{2}(1+e_p), &
 S_2 &= \frac{2\bar v_s^{2}}{v_{p,2}^{2}} = 2\chi^{2}(1-e_p), \\
 T_1 &= \frac{2\bar v_s^{2}}{v_{s,1}^{2}} = \frac{2}{1-e_s}, &
 T_2 &= \frac{2\bar v_s^{2}}{v_{s,2}^{2}} = \frac{2}{1+e_s},
\end{align}
\new{and the dimensionless ray parameter $q=2\bar v_s^{2}\,p=S_1\sin^2\theta_1$, the classical coefficients become}
\begin{equation}\label{eq:app_abcd}
\left\{\begin{array}{ll}
 a = 2\bar\rho\,A, \quad A \equiv e_r - e\,q^{2}, \quad e=e_s+e_r, \\[0.3em]
 b = \bar\rho\,B, \quad B \equiv 1-2e\,q^{2}+e_r, \\[0.3em]
 c = \bar\rho\,C, \quad C \equiv 1+2e\,q^{2}-e_r, \\[0.3em]
 d = 4\bar\rho\,\bar v_s^{2}\,e,
\end{array}\right.
\end{equation}
\new{while the slowness terms are}
\begin{equation}\label{eq:app_MN}
 M_1^{2} = 2\bar v_s^{2}\frac{\cos^{2}\theta_1}{v_{p,1}^{2}} = S_1-q^{2}, \quad
 M_2^{2} = 2\bar v_s^{2}\frac{\cos^{2}\theta_2}{v_{p,2}^{2}} = S_2-q^{2},
\end{equation}
\begin{equation}
 N_1^{2} = 2\bar v_s^{2}\frac{\cos^{2}\phi_1}{v_{s,1}^{2}} = T_1-q^{2}, \quad
 N_2^{2} = 2\bar v_s^{2}\frac{\cos^{2}\phi_2}{v_{s,2}^{2}} = T_2-q^{2}.
\end{equation}
\new{Substituting into $\mathcal{E}$, $F$, $\mathcal{G}$, and $H$ yields}
\begin{equation}\label{eq:app_EF}
 \mathcal{E} = \frac{\bar\rho}{\bar v_s\sqrt{2}}(BM_1+\varepsilon CM_2), \quad
 F = \frac{\bar\rho}{\bar v_s\sqrt{2}}(BN_1+CN_2),
\end{equation}
\begin{equation}\label{eq:app_GH}
 \mathcal{G} = 2\bar\rho(A-\varepsilon e M_1N_2), \quad
 H = 2\bar\rho(A-e M_2N_1).
\end{equation}
\new{The identity}
\begin{equation}\label{eq:app_BC_identity}
 BC - 4e\,q^{2}A = 1-e_r^{2}
\end{equation}
\new{allows $D_\varepsilon$ to be written without explicit dependence on $\bar\rho$ or $\bar v_s$ separately:}
\begin{equation}\label{eq:app_Deps}
 D_\varepsilon = \frac{\bar\rho^{2}}{2\bar v_s^{2}}(P+\varepsilon Q),
\end{equation}
\new{with}
\begin{align}\label{eq:app_PQ_chavent}
 P &= M_1\left(B^{2}N_1 + (1-e_r^{2})N_2\right) + 4e^{2}q^{2}M_1M_2N_1N_2, \\
 Q &= M_2\left(C^{2}N_2 + (1-e_r^{2})N_1\right) + 4q^{2}A^{2},
\end{align}
\new{so that}
\begin{equation}\label{eq:app_Rpp_final}
 R_{PP} = \frac{P-Q}{P+Q}.
\end{equation}
\new{Equations~}\eqref{eq:app_PQ_chavent}\new{--}\eqref{eq:app_Rpp_final} \new{ are algebraically equivalent to the classical Knott--Zoeppritz system~}\eqref{eq:app_Rpp_eps} \new{ when contrasts and background parameters are defined as in }\eqref{eq:app_contrasts}\new{--}\eqref{eq:app_background}\new{. Two useful properties of this parameterization are that (i)~$R_{PP}$ is independent of $\bar\rho$, and (ii)~background velocities enter only through $\chi=\bar v_s/\bar v_p$, reducing the local parameter count from six to four.}
\paragraph{Mapping to the multilayer notation}
\new{The forward model in the \emph{Discretized forward modelling} subsection uses the equivalent parameterization}
\begin{equation}\label{eq:app_chi_def}
 \chi = \frac{v_{s,2}^{2}+v_{s,1}^{2}}{v_{p,2}^{2}+v_{p,1}^{2}}
 = \frac{2\bar v_s^{2}}{\frac{2v_{p,1}^{2}v_{p,2}^{2}}{v_{p,1}^{2}+v_{p,2}^{2}}}
 = \frac{\bar v_s^{2}}{\bar v_p^{2}},
\end{equation}
\new{with multiplicative relations $(1-e_p)S_1=\chi$, $(1+e_p)S_2=\chi$, $(1-e_s)T_1=1$, and $(1+e_s)T_2=1$. Defining $q^2=S_1\sin^2\theta_1$, $f=1-e_r^2$, $D=e\,q^2$, $A=e_r-2D$, $K=2D-A$, $B=1-K$, and $C=1+K$, equations~}\eqref{eq:app_PQ_chavent}\new{--}\eqref{eq:app_Rpp_final} \new{become}
\begin{align}
 P &= M_1\left(B^{2}N_1 + fN_2\right) + 16eD\,M_1M_2N_1N_2, \\
 Q &= M_2\left(C^{2}N_2 + fN_1\right) + 4q^{2}A^{2}, \\
 R_{PP} &= \frac{P-Q}{P+Q},
\end{align}
\new{which is the sequence implemented in equations~}\eqref{eq:interface_params}\new{--}\eqref{eq:reflection_coeff} \new{and in the adjoint recursion of Appendix~}\ref{app:adjoint_details}\new{.} \new{The factors of two in $A$, $K$, and the $16eD$ coupling term are consistent with the rescaled $S_i$, $T_i$, and $q$ definitions above; numerical tests confirm agreement between the classical and reformulated expressions to machine precision for representative models and incidence angles.}



\segappend{Adjoint-state derivation details}{app:adjoint_details}

Suppose that we have a scalar functional for each interface $(i,j)$:

\new{Lagrangian displayed in smaller type for readability.}
{\scriptsize
\begin{align}
 L &= R_{ij} \notag\\
 &\quad + (e_{ij} - e_{s,ij} - e_{r,ij})\,\lambda_{1} \notag\\
 &\quad + (f_{ij} - 1 + e_{r,ij}^2)\,\lambda_{2} \notag\\
 &\quad + \big(S_{1,ij}(1 - e_{p,ij}) - \chi_{ij}\big)\,\lambda_{3} \notag\\
 &\quad + \big(S_{2,ij}(1 + e_{p,ij}) - \chi_{ij}\big)\,\lambda_{4} \notag\\
 &\quad + \big(T_{1,ij}(1 - e_{s,ij}) - 1\big)\,\lambda_{5} \notag\\
 &\quad + \big(T_{2,ij}(1 + e_{s,ij}) - 1\big)\,\lambda_{6} \notag\\
 &\quad + \big(q^2_{ij} - S_{1,ij} \sin^2 \theta_{j}\big)\,\lambda_{7} \notag\\
 &\quad + \big(M_{1,ij} - \sqrt{S_{1,ij} - q^2_{ij}}\big)\,\lambda_{8} \notag\\
 &\quad + \big(M_{2,ij} - \sqrt{S_{2,ij} - q^2_{ij}}\big)\,\lambda_{9} \notag\\
 &\quad + \big(N_{1,ij} - \sqrt{T_{1,ij} - q^2_{ij}}\big)\,\lambda_{10} \notag\\
 &\quad + \big(N_{2,ij} - \sqrt{T_{2,ij} - q^2_{ij}}\big)\,\lambda_{11} \notag\\
 &\quad + (D_{ij} - e_{ij} q^2_{ij})\,\lambda_{12} \notag\\
 &\quad + (A_{ij} - e_{r,ij} + 2D_{ij})\,\lambda_{13} \notag\\
 &\quad + (K_{ij} - 2D_{ij} + A_{ij})\,\lambda_{14} \notag\\
 &\quad + (B_{ij} - 1 + K_{ij})\,\lambda_{15} \notag\\
 &\quad + (C_{ij} - 1 - K_{ij})\,\lambda_{16} \notag\\
 &\quad + \Big(
      P_{ij}
      - M_{1,ij}\big(B_{ij}^2 N_{1,ij} + f_{ij} N_{2,ij}\big)
      - 16 e_{ij} D_{ij} M_{1,ij} M_{2,ij} N_{1,ij} N_{2,ij}
    \Big)\,\lambda_{17} \notag\\
 &\quad + \Big(
      Q_{ij}
      - M_{2,ij}\big(C_{ij}^2 N_{2,ij} + f_{ij} N_{1,ij}\big)
      - 4 q^2_{ij} A_{ij}^2
    \Big)\,\lambda_{18} \notag\\
    &\quad + \big(R_{ij} - \frac{P_{ij} - Q_{ij}}{P_{ij} + Q_{ij}}\big)\,\lambda_{19} .
\end{align}

The adjoint variables are determined by requiring the stationarity of $L$ with respect to all intermediate states, which yields the backward recursion (reverse sweep) given by
\begin{align}\label{eq:reverse_sweep_start}
 \lambda_{19} &= -\frac{1}{P_{ij} + Q_{ij}}, \\
 \lambda_{18} &= -(1 + R_{ij})\lambda_{19}, \\
 \lambda_{17} &= -(-1 + R_{ij})\lambda_{19}, \\
 \lambda_{16} &= 2 M_{2,ij} C_{ij} N_{2,ij}\,\lambda_{18}, \\
 \lambda_{15} &= 2 M_{1,ij} B_{ij} N_{1,ij}\,\lambda_{17}, \\
 \lambda_{14} &= \lambda_{16} - \lambda_{15}, \\
 \lambda_{13} &= -\lambda_{14} + 8 q^2_{ij} A_{ij}\,\lambda_{18}, \\
 \lambda_{12} &= -2\lambda_{13} + 2\lambda_{14}
 + 16 e_{ij} M_{1,ij} M_{2,ij} N_{1,ij} N_{2,ij}\,\lambda_{17}, \\
 \lambda_{11} &=
 \frac{M_{1,ij} f_{ij}\,\lambda_{17}
 + 16 e_{ij} D_{ij} M_{1,ij} M_{2,ij} N_{1,ij}\,\lambda_{17}
 + M_{2,ij} C_{ij}^2\,\lambda_{18}}{2 N_{2,ij}}, \\
 \lambda_{10} &=
 \frac{M_{1,ij} B_{ij}^2\,\lambda_{17}
 + 16 e_{ij} D_{ij} M_{1,ij} M_{2,ij} N_{2,ij}\,\lambda_{17}
 + M_{2,ij} f_{ij}\,\lambda_{18}}{2 N_{1,ij}}, \\
 \lambda_{9} &=
 \frac{16 e_{ij} D_{ij} M_{1,ij} N_{1,ij} N_{2,ij}\,\lambda_{17}
 + (C_{ij}^2 N_{2,ij} + f_{ij} N_{1,ij})\,\lambda_{18}}{2 M_{2,ij}}, \\
 \lambda_{8} &=
 \frac{16 e_{ij} D_{ij} M_{2,ij} N_{1,ij} N_{2,ij}\,\lambda_{17}
 + (B_{ij}^2 N_{1,ij} + f_{ij} N_{2,ij})\,\lambda_{17}}{2 M_{1,ij}}, \\
 \lambda_{7} &= - \lambda_{8} - \lambda_{9} - \lambda_{10} - \lambda_{11}
 + e_{ij}\,\lambda_{12} + 4 A_{ij}^2\,\lambda_{18}, \\
 \lambda_{6} &= \frac{\lambda_{11}}{1 + e_{s,ij}}, \\
 \lambda_{5} &= \frac{\lambda_{10}}{e_{s,ij} - 1}, \\
 \lambda_{4} &= \frac{\lambda_{9}}{1 + e_{p,ij}}, \\
 \lambda_{3} &= \frac{\lambda_{7}\sin^2 \theta_{j} + \lambda_{8}}{1 - e_{p,ij}}, \\
 \lambda_{2} &= M_{1,ij} N_{2,ij}\,\lambda_{17} + M_{2,ij} N_{1,ij}\,\lambda_{18}, \\
 \lambda_{1} &= q^2_{ij}\,\lambda_{12} + 16 D_{ij} M_{1,ij} M_{2,ij} N_{1,ij} N_{2,ij}\,\lambda_{17}.\label{eq:reverse_sweep_end}
\end{align}
}
The final derivatives with respect to $(e_{r,ij}, e_{p,ij}, e_{s,ij}, \chi_{ij})$ are then calculated by direct differentiation of the forward equations:
\begin{align}
 \frac{\partial L}{\partial e_{r,ij}} = \frac{\partial R_{ij}}{\partial e_{r,ij}}
 &= -\lambda_{1} + 2 \lambda_{2} e_{r,ij} - \lambda_{13}, \\
 \frac{\partial L}{\partial e_{p,ij}} = \frac{\partial R_{ij}}{\partial e_{p,ij}}
 &= \lambda_{4}S_{2,ij} - \lambda_{3}S_{1,ij}, \\
 \frac{\partial L}{\partial e_{s,ij}} = \frac{\partial R_{ij}}{\partial e_{s,ij}}
 &= -\lambda_{1} - \lambda_{5} T_{1,ij} + \lambda_{6} T_{2,ij}, \\
 \frac{\partial L}{\partial \chi_{ij}} = \frac{\partial R_{ij}}{\partial \chi_{ij}}
 &= -\lambda_{3} - \lambda_{4}.
\end{align}

\segappend{Expanded gradients with respect to physical parameters}{app:expanded_gradients}

To obtain gradients with respect to the physical parameters, we apply the chain rule through the definitions of $e_{r,ij}$, $e_{p,ij}$, $e_{s,ij}$ and $\chi_{ij}$. Each parameter at depth index $i$ influences at most two adjacent interfaces, giving rise to two contributions that must be summed.

\subsection{Density}

For density, the chain rule is applied via
\begin{align}
 e_{r,ij} &= \frac{\rho_{i+1} - \rho_i}{\rho_{i+1} + \rho_i}.
\end{align}
The gradient with respect to $\rho_i$ has two contributions: one from interface $i$ and one from interface $i-1$ (where $\rho_i$ appears as the density below that interface). The contribution from interface $i$ is
\begin{equation}
 \left(\frac{\partial J}{\partial \rho_i}\right)_{\text{interface }i} =
 \left(
 \frac{\rho_i - \rho_{i+1}}{(\rho_i + \rho_{i+1})^2}
 - \frac{1}{\rho_i + \rho_{i+1}}
 \right) \frac{\partial J}{\partial R_{ij}}\frac{\partial R_{ij}}{\partial e_{r,ij}}.
\end{equation}
The contribution from interface $i-1$ (where $\rho_i$ is the density below the interface) is
\begin{equation}
 \left(\frac{\partial J}{\partial \rho_i}\right)_{\text{interface }i-1} =
 \left(
 \frac{\rho_{i-1} - \rho_i}{(\rho_{i-1} + \rho_i)^2}
 + \frac{1}{\rho_{i-1} + \rho_i}
 \right) \frac{\partial J}{\partial R_{i-1,j}}\frac{\partial R_{i-1,j}}{\partial e_{r,i-1,j}}.
\end{equation}
The total gradient is the sum of these two contributions:
\begin{equation}
 \frac{\partial J}{\partial \rho_i} = \left(\frac{\partial J}{\partial \rho_i}\right)_{\text{interface }i} + \left(\frac{\partial J}{\partial \rho_i}\right)_{\text{interface }i-1}.
\end{equation}

\subsection{P-wave velocity}


The P-wave velocity enters through both $e_{p,ij}$ and $\chi_{ij}$. The gradient with respect to $v_{p,i}$ has two contributions. The contribution from interface $i$ is
\begin{equation}
 \left(\frac{\partial J}{\partial v_{p,i}}\right)_{\text{interface }i} =
 -\frac{4 v_{p,i}v_{p,i+1}^{2}}{(v_{p,i}^{2}+v_{p,i+1}^{2})^{2}}
 \frac{\partial J}{\partial R_{ij}}\frac{\partial R_{ij}}{\partial e_{p,ij}}
 -\frac{2 v_{p,i}(v_{s,i}^{2}+v_{s,i+1}^{2})}{(v_{p,i}^{2}+v_{p,i+1}^{2})^{2}}
 \frac{\partial J}{\partial R_{ij}}\frac{\partial R_{ij}}{\partial \chi_{ij}}.
\end{equation}
The contribution from interface $i-1$ (where $v_{p,i}$ is the P-wave velocity below the interface) is
\begin{align}
 \left(\frac{\partial J}{\partial v_{p,i}}\right)_{\text{interface }i-1} &=
 \frac{4 v_{p,i}v_{p,i-1}^{2}}{(v_{p,i-1}^{2}+v_{p,i}^{2})^{2}}
 \frac{\partial J}{\partial R_{i-1,j}}\frac{\partial R_{i-1,j}}{\partial e_{p,i-1,j}} \\
 &\quad
 -\frac{2 v_{p,i}(v_{s,i-1}^{2}+v_{s,i}^{2})}{(v_{p,i-1}^{2}+v_{p,i}^{2})^{2}}
 \frac{\partial J}{\partial R_{i-1,j}}\frac{\partial R_{i-1,j}}{\partial \chi_{i-1,j}}.
\end{align}
The total gradient is the sum of these two contributions:
\begin{equation}
 \frac{\partial J}{\partial v_{p,i}} = \left(\frac{\partial J}{\partial v_{p,i}}\right)_{\text{interface }i} + \left(\frac{\partial J}{\partial v_{p,i}}\right)_{\text{interface }i-1}.
\end{equation}

\subsection{S-wave velocity}

Similarly, the S-wave velocity appears in $e_{s,ij}$ and $\chi_{ij}$. The gradient with respect to $v_{s,i}$ has two contributions. The contribution from interface $i$ is
\begin{equation}
 \left(\frac{\partial J}{\partial v_{s,i}}\right)_{\text{interface }i} =
 -\frac{4 v_{s,i}v_{s,i+1}^{2}}{(v_{s,i}^{2}+v_{s,i+1}^{2})^{2}}
 \frac{\partial J}{\partial R_{ij}}\frac{\partial R_{ij}}{\partial e_{s,ij}}
 +\frac{2 v_{s,i}}{v_{p,i}^{2}+v_{p,i+1}^{2}}
 \frac{\partial J}{\partial R_{ij}}\frac{\partial R_{ij}}{\partial \chi_{ij}}.
\end{equation}
The contribution from interface $i-1$ (where $v_{s,i}$ is the S-wave velocity below the interface) is
\begin{align}
 \left(\frac{\partial J}{\partial v_{s,i}}\right)_{\text{interface }i-1} &=
 \frac{4 v_{s,i}v_{s,i-1}^{2}}{(v_{s,i-1}^{2}+v_{s,i}^{2})^{2}}
 \frac{\partial J}{\partial R_{i-1,j}}\frac{\partial R_{i-1,j}}{\partial e_{s,i-1,j}} \\
 &\quad
 +\frac{2 v_{s,i}}{v_{p,i-1}^{2}+v_{p,i}^{2}}
 \frac{\partial J}{\partial R_{i-1,j}}\frac{\partial R_{i-1,j}}{\partial \chi_{i-1,j}}.
\end{align}
The total gradient is the sum of these two contributions:
\begin{equation}
 \frac{\partial J}{\partial v_{s,i}} = \left(\frac{\partial J}{\partial v_{s,i}}\right)_{\text{interface }i} + \left(\frac{\partial J}{\partial v_{s,i}}\right)_{\text{interface }i-1}.
\end{equation}

\subsection{Assembly of depth-indexed gradients}

For each physical parameter at depth index $i$, the gradient contributions from the two adjacent interfaces are computed as shown above. Since the physical parameters do not depend on angle, the final gradient with respect to each parameter is obtained by summing over all angles $j$:
\begin{align}
 \frac{\partial J}{\partial \rho_i} &= \sum_{j=0}^{N_a-1} \left[\left(\frac{\partial J}{\partial \rho_i}\right)_{\text{interface }i} + \left(\frac{\partial J}{\partial \rho_i}\right)_{\text{interface }i-1}\right], \\
 \frac{\partial J}{\partial v_{p,i}} &= \sum_{j=0}^{N_a-1} \left[\left(\frac{\partial J}{\partial v_{p,i}}\right)_{\text{interface }i} + \left(\frac{\partial J}{\partial v_{p,i}}\right)_{\text{interface }i-1}\right], \\
 \frac{\partial J}{\partial v_{s,i}} &= \sum_{j=0}^{N_a-1} \left[\left(\frac{\partial J}{\partial v_{s,i}}\right)_{\text{interface }i} + \left(\frac{\partial J}{\partial v_{s,i}}\right)_{\text{interface }i-1}\right].
\end{align}
These gradients, collected into $\partial J_{\mathrm{data}}/\partial v_{p,i}$, $\partial J_{\mathrm{data}}/\partial v_{s,i}$, and $\partial J_{\mathrm{data}}/\partial \rho_i$ for all depth indices $i$, are combined with regularization and chain-rule terms to form $\partial J_{\mathrm{tot}}/\partial \mathbf{z}$ for equation~\eqref{eqn:5}.

\end{document}